\documentclass{article}

\usepackage{arxiv}

\usepackage[utf8]{inputenc} 
\usepackage[T1]{fontenc}    
\usepackage{hyperref}       
\usepackage{url}            
\usepackage{booktabs}       
\usepackage{amsfonts}       
\usepackage{nicefrac}       
\usepackage{microtype}      
\usepackage{lipsum}
\usepackage{graphicx}
\graphicspath{ {./images/} }
\usepackage{amsmath,amsfonts}
\usepackage{algorithm}
\usepackage{array}
\usepackage[caption=false,font=normalsize,labelfont=sf,textfont=sf]{subfig}
\usepackage{textcomp}
\usepackage{stfloats}
\usepackage{url}
\usepackage{multirow}

\usepackage{verbatim}
\usepackage{graphicx}
\usepackage{cite}
\usepackage{algpseudocode}

\newtheorem{definition}{Definition}
\newtheorem{theorem}{Theorem}

\newtheorem{proofthm}{Proof}
\usepackage{graphicx}
\usepackage{newtxmath}
\usepackage{subcaption}
\usepackage{orcidlink}

\title{Privacy-Preserving Decentralized Federated Learning via Explainable Adaptive Differential Privacy}

\author{
 Fardin Jalil Piran \\
 School of Mechanical, Aerospace, \\ and Manufacturing Engineering\\
 University of Connecticut\\
 Storrs, CT 06269 \\
  \texttt{fardin.jalil\_piran@uconn.edu} \\
   \And
 Zhiling Chen \\
 School of Mechanical, Aerospace, \\ and Manufacturing Engineering\\
 University of Connecticut\\
 Storrs, CT 06269 \\
  \texttt{zhiling.chen@uconn.edu} \\
   \And
 Yang Zhang \\
 School of Mechanical, Aerospace, \\ and Manufacturing Engineering\\
 University of Connecticut\\
 Storrs, CT 06269 \\
  \texttt{yang.3.zhang@uconn.edu} \\
   \And
 Qianyu Zhou \\
 School of Mechanical, Aerospace, \\ and Manufacturing Engineering\\
 University of Connecticut\\
 Storrs, CT 06269 \\
  \texttt{qianyu.zhou@uconn.edu} \\
   \And
 Jiong Tang \\
 School of Mechanical, Aerospace, \\ and Manufacturing Engineering\\
 University of Connecticut\\
 Storrs, CT 06269 \\
  \texttt{jiong.tang@uconn.edu} \\
   \And
 Farhad Imani \\
 School of Mechanical, Aerospace, \\ and Manufacturing Engineering\\
 University of Connecticut\\
 Storrs, CT 06269 \\
  \texttt{farhad.imani@uconn.edu} \\
}

\begin{document}
\maketitle

\begin{abstract}
Decentralized federated learning enables collaborative model training without a central server, but shared model updates can still leak sensitive information through inversion, reconstruction, and membership inference attacks. Differential privacy offers formal protection, yet existing decentralized methods operate without visibility into the noise already injected by previous participants. Each client therefore adds a full, worst-case perturbation at every step, and the accumulated noise degrades accuracy well below what the privacy requirement actually demands. We present PrivateDFL, a decentralized and privacy-preserving framework that pairs hyperdimensional computing with a transparent noise accountant. The accountant tracks the cumulative perturbation present in the shared model and lets each client add only the minimal incremental noise needed to satisfy its privacy budget. We prove that every transmitted model satisfies the target privacy guarantee, and that under this accounting the cumulative noise grows only logarithmically in the number of clients and rounds, rather than the far faster super-linear growth incurred without accounting. This yields a substantially tighter balance between privacy and accuracy than prior approaches. Across image, speech, and wearable-sensor benchmarks, and under both identically and non-identically distributed data, PrivateDFL surpasses centrally trained Transformer-based and deep neural network baselines, improving accuracy by 16 percent on images, 62 percent on speech, and 14 percent on wearable sensing over the strongest baseline in each case, while reducing inference latency by up to 119 times and energy consumption by up to 143 times. These properties make PrivateDFL a practical solution for privacy-preserving collaborative learning in settings where sensitive data cannot be centralized, such as healthcare and human-activity monitoring.
\end{abstract}

\keywords{Decentralized Federated Learning \and
Privacy-Preserving Machine Learning \and Explainable Artificial Intelligence \and
Differential Privacy \and Hyperdimensional Computing}

\section{Introduction}
\begin{figure}[!t]
\centering
\subfloat[]{\includegraphics[width=0.37\textwidth]{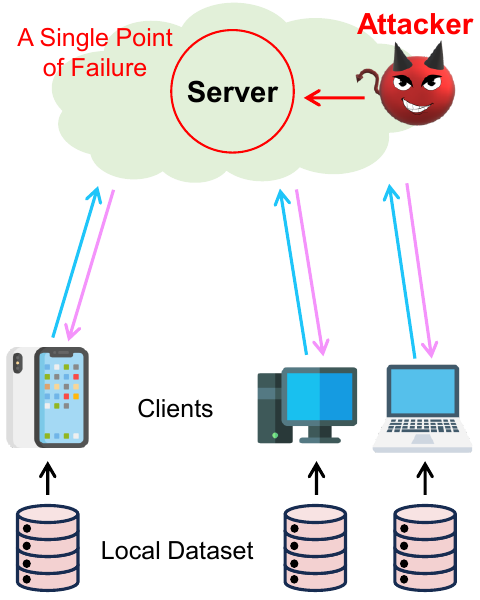}%
\label{fig:CFL}}
\hspace{0.02\textwidth}%
\subfloat[]{\includegraphics[width=0.37\textwidth]{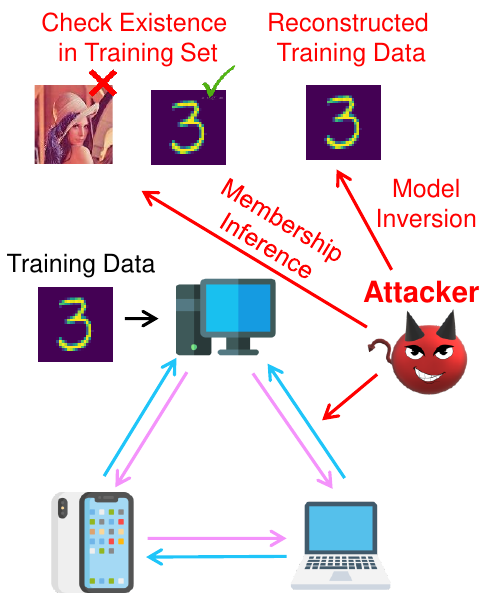}%
\label{fig:DFL}}
\caption{(a) Centralized federated learning framework. (b) Decentralized federated learning setting showing potential data leakage through model inversion and membership inference attacks.}
\label{fig:CFLandDFL}
\end{figure}

Federated Learning (FL) has emerged as a powerful paradigm for collaborative model training without the need to centralize raw data, thereby reducing communication overhead and limiting direct exposure of sensitive information such as financial and healthcare records~\cite{shirvani2023survey,jibinsingh2025fl}. In its conventional form, Centralized Federated Learning (CFL) employs a single server to aggregate local model updates from distributed clients, as illustrated in Fig.~\ref{fig:CFL}. Although CFL mitigates data-sharing concerns, the reliance on a central coordinator introduces a single point of failure and heightens susceptibility to privacy and security threats, including model leakage, reconstruction of sensitive features, and gradient-based inference attacks~\cite{rodriguez2023survey}.

To overcome these limitations, Decentralized Federated Learning (DFL) removes the central coordinator by enabling direct peer-to-peer communication among clients~\cite{yuan2024decentralized,li2025robust}. This architecture enhances scalability, robustness, and fault tolerance, and is well-suited for distributed learning in sensor-rich environments such as smart manufacturing, autonomous systems, and Internet of Things (IoT) networks~\cite{liu2024unraveling,bhattacharya2022review}. Despite these advantages, DFL remains vulnerable to sophisticated privacy attacks. Model inversion attacks reconstruct sensitive features of the training data from intermediate model updates~\cite{fredrikson2015model}, and membership inference attacks determine whether specific data instances contributed to training, as illustrated in Fig.~\ref{fig:DFL}~\cite{shokri2017membership}. These vulnerabilities underscore the need for secure and trustworthy decentralized learning frameworks that protect sensitive information while preserving high model utility.

A variety of privacy-enhancing techniques have been explored to mitigate these vulnerabilities. Early approaches rely on anonymization, which removes or obscures identifiable attributes in shared model updates. Anonymization alone, however, provides limited protection against modern inference attacks, since adversaries can still reconstruct sensitive information from gradients or output statistics even when explicit identifiers are removed~\cite{fredrikson2015model}. This weakness is especially pronounced in high-dimensional models, where subtle statistical correlations across many parameters can inadvertently leak private information.

More rigorous cryptographic solutions, such as Homomorphic Encryption (HE), enable computation directly on encrypted data and offer strong privacy protection~\cite{aono2017privacy,qin2025practical}. HE, however, incurs substantial computational and memory overhead, rendering it impractical for large-scale decentralized learning on resource-constrained clients~\cite{zhu20182p}. Secure Multiparty Computation (SMC) allows clients to jointly compute functions without revealing their local data, but its heavy communication and strict synchronization requirements limit its feasibility in real-world DFL environments with heterogeneous devices and unreliable networks. These challenges emphasize the need for lightweight, scalable, and flexible privacy-preserving mechanisms for decentralized settings.

\begin{figure}[!t]
\centering
\includegraphics[width=0.66\textwidth]{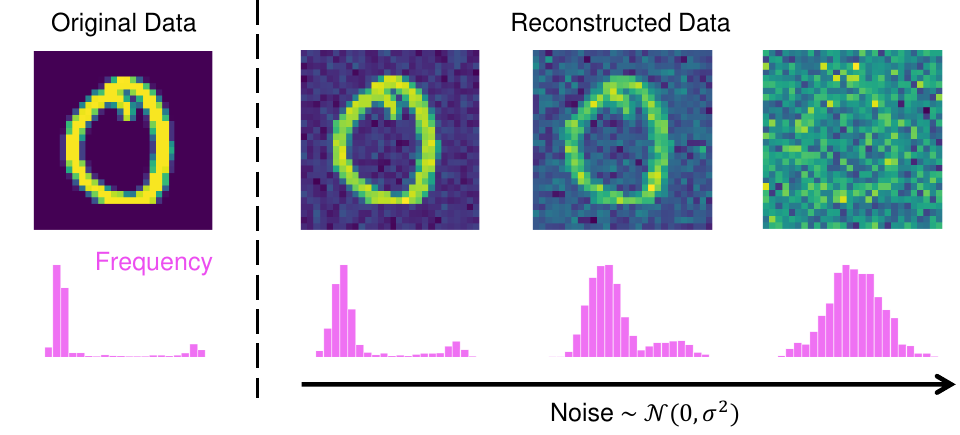}
\caption{Reconstruction of a differentially protected data point under varying noise levels.}
\label{fig:NoisySample}
\end{figure}

Differential Privacy (DP) has become a widely adopted mechanism for mitigating privacy risks in DFL. By injecting carefully calibrated noise into model updates, DP reduces the influence of individual data points and protects against attacks such as model inversion and membership inference~\cite{erlingsson2014rappor,byun2025improving}. Achieving an appropriate balance between privacy and utility remains challenging: excessive noise severely degrades accuracy, whereas insufficient noise leaves the system exposed to leakage~\cite{kim2021federated,ren2022grnn}. As illustrated in Fig.~\ref{fig:NoisySample}, the level of injected noise directly affects both the strength of the privacy guarantee and the resulting model performance. These considerations motivate adaptive, context-aware noise mechanisms that preserve formal privacy guarantees without compromising accuracy.

Standard DFL operates in a black-box manner, giving clients no visibility into the privacy noise already injected by previous participants. As a result, existing DP-based DFL methods assume worst-case exposure and add a full dose of DP noise at every update, even when substantial noise has already accumulated in the shared model. This redundant injection inflates variance, accelerates accuracy degradation, and undermines effective privacy--utility tradeoffs during long training sequences. The absence of mechanisms for inspecting how noise evolves across communication rounds further renders privacy guarantees opaque and prevents meaningful auditing in safety-critical applications~\cite{khalid2023privacy}. A further gap is that prior decentralized DP methods provide no formal account of how the guarantee holds against an adversary who observes every model exchanged along the network, leaving both the privacy analysis and the privacy--utility balance incompletely characterized. These limitations motivate a DFL framework that tracks cumulative noise, prevents unnecessary amplification, and provides transparent, verifiable enforcement of privacy budgets throughout collaborative training.

\begin{figure}[!t]
\centering
\includegraphics[width=0.8\textwidth]{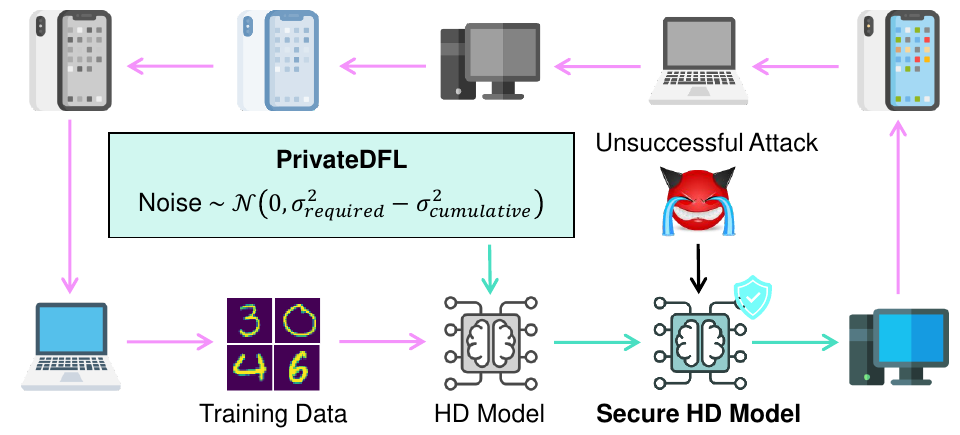}
\caption{Overview of the PrivateDFL framework with XAI-guided noise accounting and decentralized HD-based model updates.}
\label{fig:DP_overview}
\end{figure}

To address these gaps, we introduce Private Decentralized Federated Learning (PrivateDFL), a fully decentralized and explainable DP framework. As illustrated in Fig.~\ref{fig:DP_overview}, PrivateDFL incorporates an eXplainable Artificial Intelligence (XAI)-guided noise accountant that records the cumulative noise injected by all previous clients and rounds, computes the noise required to satisfy the current DP budget after each local update, and injects only the incremental difference between these two quantities. Here, the explainability is intrinsic rather than borrowed from an external method: the HyperDimensional computing (HD) classifier exposes an interpretable structure of class prototypes with explicit update rules, and the accountant makes the injected privacy noise transparent, traceable, and mathematically verifiable across decentralized exchanges rather than acting as a post-hoc explainer. PrivateDFL also integrates a HD model whose lightweight vector operations support fast, low-energy updates on resource-constrained devices. Clients are organized in a ring topology and sequentially update a shared HD model using their local data, enabling a fully serverless and communication-efficient training pipeline. Because the accountant injects only the minimal incremental noise required at each step, PrivateDFL preserves accuracy even under strict privacy budgets, yielding a favorable and interpretable privacy--utility balance. Our main contributions are summarized as follows.
\begin{itemize}
    \item We present PrivateDFL, the first DFL framework that unifies DP with an XAI-guided noise accountant. The framework provides transparent and auditable tracking of privacy noise throughout all training rounds, together with a formal guarantee that every model exchanged between clients satisfies the target $(\epsilon,\delta)$ budget against an adversary that observes every exchange.
    \item We propose an adaptive privacy mechanism that monitors the cumulative noise injected by previous clients, determines the noise required to satisfy the current privacy budget, and injects only the incremental difference. We prove that the resulting cumulative noise grows only logarithmically in the number of clients and rounds, in contrast to the factorial growth incurred without noise accounting, which substantially improves the privacy--utility tradeoff in fully serverless DFL.
    \item We integrate an HD model whose lightweight, interpretable vector operations enable privacy accounting in high-dimensional feature spaces while keeping inference latency and energy consumption low, making PrivateDFL practical for resource-constrained devices.
    \item We conduct comprehensive experiments on MNIST, ISOLET, and UCI-HAR under Independent and Identically Distributed (IID) and non-IID conditions, benchmarking PrivateDFL against Transformer-based and deep learning models trained with Differential Privacy-Stochastic Gradient Descent (DP-SGD) and Rényi Differential Privacy (RDP). PrivateDFL surpasses these centralized baselines in accuracy, latency, and energy efficiency while maintaining formal $(\epsilon,\delta)$ privacy guarantees.
\end{itemize}

The remainder of this paper is organized as follows. Section~\ref{sec:Related Work} reviews prior work on DFL, DP, and HD. Section~\ref{sec:Preliminary} introduces the fundamentals of HD and DP. Section~\ref{sec:Research Methodology} presents the proposed PrivateDFL framework and its implementation. Section~\ref{sec:Experimental Design} outlines the experimental setup, and Section~\ref{sec:Experimental Results} reports and analyzes the results. Section~\ref{sec:Discussion and Limitations} discusses the findings and limitations of the framework. Finally, Section~\ref{sec:Conclusions and Future Work} summarizes the contributions and highlights future research directions.

\section{Research Background}
\label{sec:Related Work}

\subsection{Decentralized Federated Learning}
\label{sec:RW Decentralized Federated Learning}

CFL enables collaborative model training but faces well-documented limitations, including client heterogeneity~\cite{li2020federated}, constrained communication resources~\cite{fang2021over}, and the overhead of repeated aggregation~\cite{imteaj2021survey}. More critically, the central server is a single point of failure and a primary target for malicious attacks~\cite{bagdasaryan2020backdoor,tolpegin2020data}. These limitations motivate DFL, which removes the central coordinator and distributes responsibilities across clients.

Several DFL frameworks have been proposed. Fully DFL~\cite{lalitha2018fully} replaces global aggregation with direct peer-to-peer communication, and decentralized federated averaging~\cite{lian2017can} extends Federated Averaging to a topology in which clients update through local neighbors. Blockchain-based DFL provides tamper-resistant aggregation without a central authority~\cite{qu2020decentralized}, and graph-based FL formalizes client interactions as a graph supporting flexible communication~\cite{yuan2024decentralized}. Collectively, these strategies improve scalability, fault tolerance, and resilience in resource-constrained environments such as IoT networks.

However, removing the central server does not eliminate privacy risks. Clients still exchange intermediate model updates, which remain vulnerable to model inversion and membership inference attacks that reveal sensitive properties of the training data. This motivates integrating DP to provide formal privacy guarantees while preserving model utility, a central and still-open challenge for DFL.

\subsection{Differential Privacy}
\label{sec:RW Differential Privacy}

DP~\cite{dwork2014algorithmic} safeguards sensitive data by injecting calibrated noise so that the output distribution changes only marginally when a single record is added or removed, keeping individual contributions indistinguishable. This limits reconstruction and membership inference attacks and has become central to privacy-preserving machine learning. A widely used instantiation is DP-SGD, which clips per-sample gradients and injects noise into aggregated gradients, offering quantifiable guarantees while preserving utility~\cite{abadi2016deep}. RDP~\cite{mironov2017renyi} refines this by characterizing privacy loss through R\'enyi divergence, providing a flexible composition mechanism for tracking cumulative loss across iterations, at the cost of additional analytical complexity.

DP has been incorporated into many FL frameworks. Gong et al.\ combined DP with HE to protect gradients, though leakage is not fully mitigated~\cite{gong2020privacy}. Wu et al.\ added Gaussian noise to parameters in multi-task FL, but the defense remains vulnerable to reconstruction and label flipping~\cite{wu2020theoretical}. Cyffers et al.\ analyzed a relaxed local-DP approach for decentralized learning~\cite{cyffers2022privacy}, which can heighten exposure to data poisoning~\cite{el2022differential}, and Tran et al.\ proposed a decentralized protocol with a rotating master role~\cite{tran2021efficient}, though collusion can still compromise privacy~\cite{el2022differential}.

Hybrid schemes combine DP with complementary mechanisms. Zhao et al.\ shared partial gradients with Gaussian perturbations through a proxy to obscure client identity~\cite{zhao2021anonymous}, and Yin et al.\ integrated functional encryption with Bayesian DP~\cite{yin2021privacy}, though the latter relies on a trusted third party. Other work adapts DP noise through data-driven encoding~\cite{gao2024amoue}, applies random noise to anonymize sensor signals in cyber-physical systems~\cite{basak2023dppt}, or structures sensitive features before applying DP in IoT systems~\cite{ali2022privacy}. More recently, XAI has been paired with DP to improve transparency; Zheng et al.\ proposed a dynamic budget-allocation scheme that uses explanatory signals to adjust noise~\cite{zheng2025awe}. As decentralized learning expands into high-stakes domains, a key challenge remains designing adaptive mechanisms that provide formal, verifiable guarantees while retaining utility.

\begin{figure*}[!t]
    \centering
    \includegraphics[width=\textwidth]{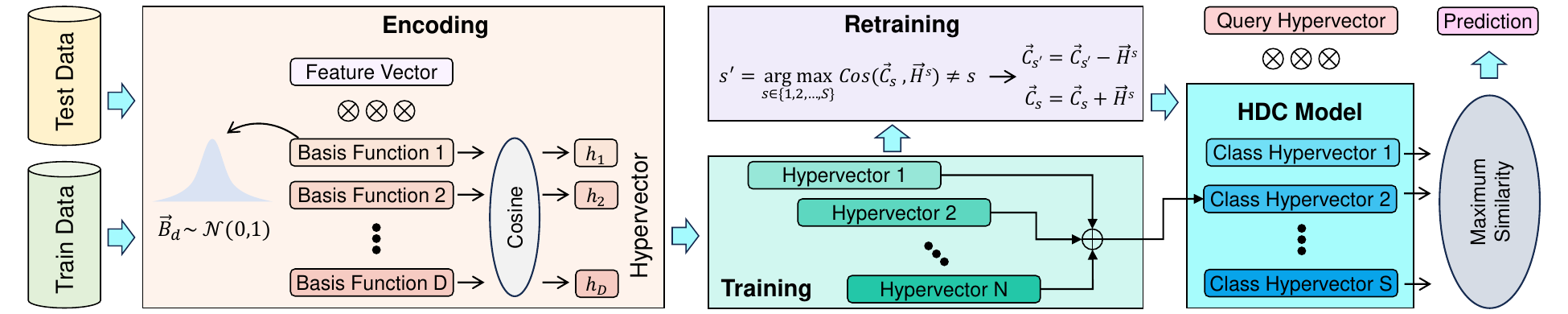}
    \caption{Overview of the hyperdimensional computing model illustrating the encoding, training, inference, and retraining stages.}
    \label{fig:HD}
\end{figure*}

\subsection{Hyperdimensional Computing}
\label{sec:RW Hyperdimensional Computing}

HD has gained attention as a lightweight and computationally efficient alternative to conventional machine learning. Its high-dimensional vector representations support simple, parallelizable operations with low inference latency and low energy cost, making HD attractive for resource-constrained and distributed environments. Recent work has begun incorporating privacy-preserving techniques into HD models. Hernandez-Cano et al.~\cite{hernandez2021prid} introduced a noise-injection strategy that masks less informative features but did not specify the noise levels required for formal guarantees. Khaleghi et al.~\cite{khaleghi2020prive} integrated DP into HD workflows without fully examining the impact on accuracy. Piran et al.~\cite{piran2025explainable} analyzed DP on HD through an explainability perspective but only for standalone systems, and subsequent work~\cite{piran2025privacy} extended DP to HD-based CFL, where reliance on a central server limited the privacy advantages and left decentralized settings unaddressed.

These studies highlight the need to advance privacy-preserving HD methods that improve transparency while maintaining strong performance in federated environments. Developing adaptive noise-injection mechanisms that support formal privacy guarantees while retaining the efficiency and interpretability of HD representations remains an important direction for trustworthy decentralized learning.

\section{Preliminary}
\label{sec:Preliminary}

\subsection{Hyperdimensional Computing}
\label{sec:RM Hyperdimensional Computing}

HD is an interpretable learning framework based on high-dimensional vector representations known as hypervectors~\cite{piran2025domain,JalilPiran2023}. Inspired by brain-like distributed computation~\cite{jalil2024privacy}, it encodes, aggregates, and compares information through simple vector operations, which yields an interpretable model with low computational cost~\cite{chen2025federated}. As shown in Fig.~\ref{fig:HD}, HD computation proceeds through four stages: encoding, training, inference, and retraining. These lightweight operations make HD well-suited for decentralized and privacy-preserving learning~\cite{piran2025privacy}.

\textbf{Encoding} maps an input feature vector $\Vec{F}$ into a $D$-dimensional hypervector $\Vec{H}$ using random basis functions $\{\Vec{B}_d\}_{d=1}^{D}$ sampled from $\mathcal{N}(0,1)$, with components defined as
\begin{equation}
h_d = \cos(\Vec{F} \cdot \Vec{B}_d).
\end{equation}
This mapping preserves feature similarity in a structured and interpretable form.

\textbf{Training} constructs class prototypes by summing hypervectors belonging to the same label. For encoded samples $\Vec{H}^s$ of class $s$, the class hypervector is
\begin{equation}
\Vec{C}_s = \sum \Vec{H}^s.
\label{eq:formClassHypervector}
\end{equation}
The resulting prototypes capture the dominant patterns shared across samples of each class.

\textbf{Inference} classifies a query hypervector $\Vec{H}^{q}$ by measuring its cosine similarity with each class prototype. The similarity is computed as
\begin{equation}
\text{Cos}(\Vec{C}_{s}, \Vec{H}^{q}) =
\frac{\Vec{C}_{s} \cdot \Vec{H}^{q}}
     {||\Vec{C}_{s}||\, ||\Vec{H}^{q}||}.
\end{equation}
The predicted label corresponds to the class with the highest similarity score:
\begin{equation}
q = \arg\max_{s} \text{Cos}(\Vec{C}_{s}, \Vec{H}^{q}).
\end{equation}
This similarity-based decision rule provides an interpretable mapping between queries and class prototypes.

\textbf{Retraining} updates class prototypes when misclassifications occur. For a misclassified sample $\Vec{H}^{s}$ predicted as $s'$, prototypes are corrected by
\begin{equation}
\Vec{C}_{s'} = \Vec{C}_{s'} - \Vec{H}^{s}, \qquad
\Vec{C}_{s}  = \Vec{C}_{s}  + \Vec{H}^{s}.
\label{eq:updateclasshv}
\end{equation}
This refinement improves accuracy and maintains consistent prototypes over successive updates.

Overall, HD computing offers interpretable and computationally efficient operations that naturally align with the requirements of decentralized learning.

\subsection{Differential Privacy}
\label{sec:P Differential Privacy}

DP provides a formal guarantee that the inclusion or removal of any single record has only a limited effect on the output of a computation. This protection is achieved by adding calibrated random noise to model updates. The privacy parameters \(\epsilon\) and \(\delta\) quantify the strength of this guarantee, where \(\epsilon>0\) denotes the privacy budget that limits how much the output distribution may change when a single record is modified, and \(\delta \in [0,1)\) is a small probability that bounds the likelihood of the guarantee failing~\cite{dwork2014algorithmic,dwork2006calibrating}.

\begin{definition}\label{def:DP}
A randomized mechanism \(\mathcal{M}\) satisfies \((\epsilon, \delta)\)-DP if, for any pair of neighboring datasets \(I_1\) and \(I_2\) differing in one record,
\begin{equation}
\Pr[\mathcal{M}(I_1)] \leq e^{\epsilon}\Pr[\mathcal{M}(I_2)] + \delta.
\end{equation}
This condition ensures that an adversary cannot reliably determine whether any specific record is present in the dataset.
\end{definition}

\begin{definition}\label{def:Gaussian Mechanism}
The Gaussian Mechanism provides \((\epsilon, \delta)\)-DP by perturbing the output of a function \(g(I)\) with Gaussian noise of variance \(\sigma_{dp}^2\):
\begin{equation}
\mathcal{M}(I) = g(I) + \mathcal{N}(0,\sigma_{dp}^2).
\label{eq:eq_dp}
\end{equation}
\end{definition}

\begin{definition}\label{def:Sensitivity}
The sensitivity \(\Delta g\) of a function \(g\) is defined as
\begin{equation}
\Delta g = \max_{I_1, I_2} \|g(I_1) - g(I_2)\|,
\label{eq:sensitive}
\end{equation}
which captures the maximum possible influence of a single record on the output.
\end{definition}

\begin{theorem}\label{thm:theorem1}
The Gaussian Mechanism satisfies \((\epsilon,\delta)\)-DP if the noise variance satisfies
\begin{equation}
\sigma_{dp}^2 \geq 2\frac{\Delta g^2}{\epsilon^2}
\ln\!\left(\frac{1.25}{\delta}\right).
\label{eq:sigma_dp}
\end{equation}
\end{theorem}

\section{Research Methodology}
\label{sec:Research Methodology}

\subsection{Threat Model}
\label{sec:RM Threat Model}

PrivateDFL targets the standard privacy setting for decentralized learning. Clients are organized in a ring and sequentially update a shared HD model, so each client passes a perturbed model to its successor without any central server. We make the following assumptions.

\emph{Adversary.} We consider an honest-but-curious adversary that observes every HD model transmitted between clients. Concretely, the adversary sees the full sequence of models exchanged along the ring across all communication rounds, and may correspond to a curious participant that inspects the models it receives, or to a passive observer of the peer-to-peer channel. The adversary follows the protocol but attempts to infer sensitive properties of any client's local training data from the models it observes, through model inversion, reconstruction, or membership inference.

\emph{Protected asset.} The asset to be protected is the contribution of each client's local training data to the shared model. Privacy is enforced at the level of individual records through the $(\epsilon,\delta)$ guarantee of Definition~\ref{def:DP}.

\emph{Guarantee.} PrivateDFL calibrates the noise so that \emph{every} model transmitted between clients satisfies $(\epsilon,\delta)$-DP for all training samples aggregated into it up to that point. Because each transmitted model is individually differentially private, the guarantee holds against an adversary that observes every exchange along the ring. This calibration depends only on the number of samples and clients aggregated so far, and is therefore identical under independently and identically distributed and non-independently distributed data partitions.

\emph{Scope.} Consistent with the honest-but-curious model, we do not address active adversaries that deviate from the protocol, including data poisoning, backdoor injection, model manipulation, or free-riding. These threats, together with collusion among participants, are outside the present scope and are discussed as directions for future work in Section~\ref{sec:Conclusions and Future Work}.

\subsection{PrivateDFL}
\label{sec:RM PrivateDFL}

PrivateDFL builds on decentralized learning, HD, and DP to provide a secure and interpretable framework for distributed model training. In this formulation, HD offers a transparent and interpretable representational structure for classification through class prototypes and explicit update rules, while DFL removes the central server and enables clients to sequentially update and exchange HD models. To protect individual data contributions during these updates, DP is incorporated with parameters \((\epsilon,\delta)\), where a smaller \(\epsilon\) indicates stronger privacy and \(\delta\) bounds the probability that the guarantee may not hold. Following standard practice, \(\delta\) is selected to scale inversely with the total number of training samples \(N\), such that \(\delta=\delta_{0}/N\) with \(\delta_{0}\in(0,1)\), ensuring that the probability of privacy leakage remains negligible as the dataset grows.

At each communication round \(r\), client \(k\) determines its DP noise contribution \(\vec{\Gamma}_{k}^{r}\) by computing the difference between the noise required to meet the privacy budget at that point, \(\vec{\xi}_{k}^{r}\), and the cumulative noise \(\vec{\Psi}_{k-1}^{r}\) already present in the received model. This incremental strategy prevents unnecessary amplification of noise and ensures that privacy guarantees are satisfied without degrading model utility.

Training begins with the collaborative construction of the HD model. The first client generates class hypervectors using Equation~\eqref{eq:formClassHypervector}, perturbs them with DP noise as described in Theorem~\ref{thm:theorem2}, and forwards the secure model to the next participant, with correctness shown in Proof~\ref{proof:proof1}. Each subsequent client receives the noisy class hypervectors, constructs its own class hypervectors from local training data, merges them with the received model, and adds additional noise according to Theorem~\ref{thm:theorem3}, justified in Proof~\ref{proof:proof2}.

\begin{theorem}\label{thm:theorem2}
The DP noise added to the HD model by the first client in the first round is
\begin{equation}
\vec{\Gamma}_{1}^{1} \sim \mathcal{N}\!\left(0, \frac{2D}{\epsilon^2} 
\ln\!\left[\frac{1.25N}{\delta_0}\right]\right).
\label{eq:theorem_2_last_equation}
\end{equation}
\end{theorem}

\begin{proofthm}\label{proof:proof1}
The required noise \(\vec{\xi}_{1}^{1}\) for the first client is obtained from Equation~\eqref{eq:sigma_dp}. 
From Equation~\eqref{eq:sensitive}, the sensitivity of the model is \(\Delta g=\sqrt{D}\), since hypervector elements lie in the bounded interval \([{-}1,1]\) (attaining \(\pm 1\) at the extrema), and the largest possible difference between two neighboring datasets yields an \(\ell_2\)-norm of \(\sqrt{D}\).

In the worst case, all \(N\) training samples for a given client contribute to a single class hypervector. 
Thus, we set \(\delta = \delta_0/N\). Substituting into Equation~\eqref{eq:sigma_dp} gives
\begin{equation}
\vec{\xi}_{1}^{1} \sim \mathcal{N}\!\left(0, \frac{2D}{\epsilon^2} 
\ln\!\left[\frac{1.25N}{\delta_0}\right]\right).
\end{equation}
Since this is the first round with no previously accumulated noise,
\begin{equation}
\vec{\Gamma}_{1}^{1} = \vec{\xi}_{1}^{1}.
\end{equation}
\end{proofthm}

\begin{theorem}\label{thm:theorem3}
For any client \(k \ge 2\) in the first round, the DP noise added to the HD model satisfies
\begin{equation}
\vec{\Gamma}_{k}^{1} \sim \mathcal{N}\!\left(0, \frac{2D}{\epsilon^{2}} \,\ln\!\left[\frac{k}{\,k-1\,}\right]\right),
\label{eq:theorem_3_last_equation}
\end{equation}
where \(k\) denotes the client index.
\end{theorem}

\begin{proofthm}\label{proof:proof2}
From Equation~\eqref{eq:sigma_dp} and the sensitivity in Equation~\eqref{eq:sensitive} with \(\Delta g=\sqrt{D}\), the cumulative noise required after client \(k\) in round~1 is
\begin{equation}
\vec{\xi}_{k}^{1} \sim \mathcal{N}\!\left(0, \frac{2D}{\epsilon^{2}} \,\ln\!\left[\frac{1.25\,k\,N}{\delta_{0}}\right]\right),
\end{equation}
using \(\delta=\delta_{0}/(kN)\) in the worst case. The model received from client \(k-1\) already contains
\begin{equation}
\vec{\Psi}_{k-1}^{1} \sim \mathcal{N}\!\left(0, \frac{2D}{\epsilon^{2}} \,\ln\!\left[\frac{1.25\,(k-1)\,N}{\delta_{0}}\right]\right).
\end{equation}

Choose \(\vec{\Gamma}_{k}^{1}\) independent of \(\vec{\Psi}_{k-1}^{1}\) so that \(\vec{\Psi}_{k-1}^{1}+\vec{\Gamma}_{k}^{1}\) attains the target variance. Then
\begin{align}
\vec{\Gamma}_{k}^{1} &\sim \mathcal{N}\!\left(0, \frac{2D}{\epsilon^{2}} \Bigl(\ln\!\left[\tfrac{1.25\,k\,N}{\delta_{0}}\right]-\ln\!\left[\tfrac{1.25\,(k-1)\,N}{\delta_{0}}\right]\Bigr)\right) \notag\\
&= \mathcal{N}\!\left(0, \frac{2D}{\epsilon^{2}} \,\ln\!\left[\frac{k}{\,k-1\,}\right]\right).
\end{align}
\end{proofthm}

Although a class prototype is a sum of many hypervectors and its magnitude grows as training proceeds, the sensitivity \(\Delta g\) measures the change induced by adding or removing a single record, not the absolute magnitude of the prototype. Adding or removing one sample alters a prototype by exactly one encoded hypervector whose entries lie in \([-1,1]\), giving an \(\ell_2\)-norm of at most \(\sqrt{D}\). The sensitivity is therefore \(\Delta g=\sqrt{D}\) regardless of how large the accumulated prototype becomes, including under the retraining rule of Equation~\eqref{eq:updateclasshv}, where a record contributes a single bounded add-and-subtract of one hypervector.

In subsequent communication rounds, each client refines the received class hypervectors using Equation~\eqref{eq:updateclasshv} with its own private training data. Although a non-independently distributed client holds samples from only a few classes, the received model already contains all \(S\) class prototypes. The client encodes each local sample, compares it against all \(S\) prototypes, and on any misclassification subtracts the hypervector from the wrongly predicted class and adds it to the true class, so a client holding few classes can still misclassify into and correct any of the \(S\) prototypes. Each client then calibrates additional DP noise by accounting for the perturbations accumulated in earlier rounds to meet the privacy constraints, and forwards the secure HD model to the next participant. The noise schedule for the first client in these rounds is stated in Theorem~\ref{thm:theorem4} and justified in Proof~\ref{proof:proof3}, while the schedule for the remaining clients is given in Theorem~\ref{thm:theorem5} and justified in Proof~\ref{proof:proof4}.

\begin{theorem}\label{thm:theorem4}
The DP noise added by the first client in round \(r\) after \(r-1\) prior communication rounds is
\begin{equation}
\vec{\Gamma}_{1}^{r} \sim \mathcal{N}\!\left(0, \frac{2D}{\epsilon^2}\,\ln\!\left[\frac{K(r-1)+1}{K(r-1)}\right]\right).
\label{eq:theorem_4_last_equation}
\end{equation}
\end{theorem}

\begin{proofthm}\label{proof:proof3}
Let \(\vec{\xi}_{1}^{r}\) denote the cumulative noise required to protect all contributions present when the first client completes round \(r\), and let \(\vec{\Psi}_{K}^{\,r-1}\) be the accumulated noise already contained in the model received from client \(K\) after \(r-1\) rounds. By Equation~\eqref{eq:sensitive}, the sensitivity remains \(\Delta g=\sqrt{D}\). In round \(r\), the model aggregates \(KN(r-1)+N\) samples, hence
\begin{equation}
\delta=\frac{\delta_0}{KN(r-1)+N}.
\end{equation}

Substituting into Equation~\eqref{eq:sigma_dp} yields
\begin{equation}
\vec{\xi}_{1}^{r} \sim \mathcal{N}\!\left(0, \frac{2D}{\epsilon^2}\,\ln\!\left[\frac{1.25\,KN(r-1)+1.25\,N}{\delta_0}\right]\right).
\end{equation}

The received model already contains
\begin{equation}
\vec{\Psi}_{K}^{\,r-1} \sim \mathcal{N}\!\left(0, \frac{2D}{\epsilon^2}\,\ln\!\left[\frac{1.25\,KN(r-1)}{\delta_0}\right]\right).
\end{equation}

Choose \(\vec{\Gamma}_{1}^{r}\) independent of \(\vec{\Psi}_{K}^{\,r-1}\) so that their sum attains the target variance. Therefore,
\begin{align}
\vec{\Gamma}_{1}^{r} &\sim \mathcal{N} \Biggl( 0,\, \frac{2D}{\epsilon^2} \biggl( \ln\!\left[\frac{1.25\,KN(r-1)+1.25\,N}{\delta_0}\right] \notag -\ln\!\left[\frac{1.25\,KN(r-1)}{\delta_0}\right] \biggr) \Biggr) \notag \\
&= \mathcal{N}\!\left( 0,\, \frac{2D}{\epsilon^2} \ln\!\left[\frac{K(r-1)+1}{K(r-1)}\right] \right).
\end{align}
\end{proofthm}

\begin{theorem}\label{thm:theorem5}
For client \(k \in \{2,\dots,K\}\) in round \(r\), after \(k-1\) clients have updated the class hypervectors, the DP noise added to the HD model is
\begin{equation}
\vec{\Gamma}_{k}^{r} \sim \mathcal{N}\!\left(0, \frac{2D}{\epsilon^{2}} \,\ln\!\left[\frac{K(r-1)+k}{K(r-1)+k-1}\right]\right).
\label{eq:theorem_5_last_equation}
\end{equation}
\end{theorem}

\begin{proofthm}\label{proof:proof4}
By Equation~\eqref{eq:sensitive}, the sensitivity is \(\Delta g=\sqrt{D}\). When client \(k\) completes round \(r\), the model aggregates \(KN(r-1)+kN\) samples, so \(\delta=\delta_{0}/(KN(r-1)+kN)\). Hence
\begin{equation}
\vec{\xi}_{k}^{r} \sim \mathcal{N}\!\left(0, \frac{2D}{\epsilon^{2}} \,\ln\!\left[\frac{1.25\,(K(r-1)+k)N}{\delta_{0}}\right]\right),
\end{equation}
and the received model already contains
\begin{equation}
\vec{\Psi}_{k-1}^{r} \sim \mathcal{N}\!\left(0, \frac{2D}{\epsilon^{2}} \,\ln\!\left[\frac{1.25\,(K(r-1)+k-1)N}{\delta_{0}}\right]\right).
\end{equation}

Choose \(\vec{\Gamma}_{k}^{r}\) independent of \(\vec{\Psi}_{k-1}^{r}\) so that their sum attains the target variance. Therefore,
\begin{align}
\vec{\Gamma}_{k}^{r} &\sim \mathcal{N}\Biggl(0,\, \frac{2D}{\epsilon^{2}} \biggl(\ln\!\left[\frac{1.25\,(K(r-1)+k)N}{\delta_{0}}\right] \notag  - \ln\!\left[\frac{1.25\,(K(r-1)+k-1)N}{\delta_{0}}\right]\biggr)\Biggr) \notag \\
&\sim \mathcal{N}\!\left(0, \frac{2D}{\epsilon^{2}} \,\ln\!\left[\frac{K(r-1)+k}{K(r-1)+k-1}\right]\right).
\end{align}
\end{proofthm}

The added noise in each communication round is determined by Theorems~\ref{thm:theorem2}--\ref{thm:theorem5}. For the first round, the noise for the initial client is given in Theorem~\ref{thm:theorem2} (Equation~\ref{eq:theorem_2_last_equation}), while the noise for clients \(k \ge 2\) is described in Theorem~\ref{thm:theorem3} (Equation~\ref{eq:theorem_3_last_equation}). In subsequent rounds (\(r \ge 2\)), the noise for the first client is defined in Theorem~\ref{thm:theorem4} (Equation~\ref{eq:theorem_4_last_equation}), and for all other clients in that round it is specified by Theorem~\ref{thm:theorem5} (Equation~\ref{eq:theorem_5_last_equation}). 

It is straightforward to observe that Equation~\ref{eq:theorem_4_last_equation} can be obtained from Equation~\ref{eq:theorem_5_last_equation} by setting \(k=1\). Similarly, Equation~\ref{eq:theorem_3_last_equation} is obtained from Equation~\ref{eq:theorem_5_last_equation} by substituting \(r=1\). Therefore, the noise injection procedure simplifies to the following rule: the first client in the first round follows Theorem~\ref{thm:theorem2}, while all other cases, including clients \(k\ge2\) in the first round and all clients in subsequent rounds, follow Theorem~\ref{thm:theorem5}. The implementation of PrivateDFL based on this unified formulation is presented in Algorithm~\ref{alg:privatedfl}.

\subsection{Cumulative Noise}
\label{sec:RM Cumulative Noise}

We now derive the total noise accumulated in PrivateDFL after any client $k$ in round $r$, and compare it with the cumulative noise that arises in a black-box setting where past perturbations cannot be tracked. This comparison highlights why tracking cumulative noise is essential for maintaining accuracy under DP. Without noise accounting, each client must inject the full required noise at every step, leading to rapid and unnecessary variance growth.

\subsubsection{Cumulative Noise Under PrivateDFL}

Let $\vec{\gamma}_{k}^{r}$ denote the cumulative noise in the HD model after client $k$ completes round $r$. Since PrivateDFL adds only the incremental noise needed to reach the target variance, we have
\begin{equation}
\vec{\gamma}_{k}^{r}
= \sum_{j=1}^{r-1}\sum_{i=1}^{K} \vec{\Gamma}_{i}^{\,j}
  + \sum_{i=1}^{k} \vec{\Gamma}_{i}^{\,r}.
\end{equation}

Using the noise schedule in Theorems~\ref{thm:theorem2}--\ref{thm:theorem5} and telescoping of the logarithmic factors, the cumulative noise simplifies to
\begin{equation}
\vec{\gamma}_{k}^{r}
\sim
\mathcal{N}\!\left(
0,\,
\frac{2D}{\epsilon^{2}}
\ln\!\left[
  \frac{1.25\, (K(r-1)+k)\, N}{\delta_{0}}
\right]
\right),
\label{eq:gamma_k_r_clean}
\end{equation}
which matches exactly the required noise $\vec{\xi}_{k}^{r}$ needed to protect all samples aggregated up to that stage:
\begin{equation}
\vec{\gamma}_{k}^{r} = \vec{\xi}_{k}^{r}.
\end{equation}

Setting $r=R$ and $k=K$ gives the final cumulative noise of the trained model:
\begin{equation}
\vec{\gamma}_{K}^{R}
\sim
\mathcal{N}\!\left(
0,\;
\frac{2D}{\epsilon^{2}}
\ln\!\left[
  \frac{1.25\, K R\, N}{\delta_{0}}
\right]
\right).
\label{eq:gamma_KR_clean}
\end{equation}
This shows that PrivateDFL exhibits only logarithmic growth in variance with respect to the number of clients and rounds.

\subsubsection{Cumulative Noise in the Black-Box Case}

In a system where noise cannot be tracked, each client must inject the full required DP noise at every round. Let $\vec{\Xi}_{k}^{r}$ denote the cumulative noise under this black-box assumption:
\begin{equation}
\vec{\Xi}_{k}^{r}
= 
\sum_{j=1}^{r}\sum_{i=1}^{k} \vec{\xi}_{i}^{\,j}.
\end{equation}

From Proof~\ref{proof:proof4},
\begin{equation}
\vec{\xi}_{i}^{j}
\sim
\mathcal{N}\!\left(
0,\;
\frac{2D}{\epsilon^{2}}
\ln\!\left[
\frac{1.25 N\, j\, i}{\delta_{0}}
\right]
\right),
\end{equation}
and summing all contributions gives
\begin{equation}
\vec{\Xi}_{k}^{r}
\sim
\mathcal{N}\!\Bigg(
0,\,
\frac{2D}{\epsilon^2}\,
\sum_{t=1}^{K(r-1)+k}
\ln\!\left(\frac{1.25Nt}{\delta_0}\right)
\Bigg).
\label{eq:Xi_kr_clean}
\end{equation}

Rewriting the sum as a logarithm of a product yields
\begin{align}
\vec{\Xi}_{k}^{r}
&\sim 
\mathcal{N}\!\Bigg(
0,\,
\frac{2D}{\epsilon^{2}}\,
\ln\!\Bigg[
\prod_{t=1}^{K(r-1)+k}
\frac{1.25Nt}{\delta_0}
\Bigg]
\Bigg)
\notag \\[4pt]
&=
\mathcal{N}\Bigg(
0,\,
\frac{2D}{\epsilon^{2}}\,
\ln\!\Bigg[
\left(\frac{1.25N}{\delta_0}\right)^{K(r-1)+k}
\Bigg] \notag \\
&\qquad\qquad
+\frac{2D}{\epsilon^{2}}\,
\ln\!\Big[(K(r-1)+k)!\Big]
\Bigg).
\label{eq:Xi_product_clean}
\end{align}

For the final model ($k=K$, $r=R$), we obtain
\begin{equation}
\vec{\Xi}_{K}^{R}
\sim
\mathcal{N}\!\left(
0,\,
\frac{2D}{\epsilon^{2}}\,
\ln\!\left[
\left(\frac{1.25N}{\delta_0}\right)^{KR}
(KR)!
\right]
\right).
\label{eq:Xi_KR_clean}
\end{equation}

The argument of the logarithm scales super-exponentially with $KR$ due to the product term $\left(\frac{1.25N}{\delta_0}\right)^{KR} (KR)!$. Applying Stirling's approximation, $\ln (KR)! = KR\ln(KR) - KR + \mathcal{O}(\ln KR)$, the resulting variance scales as $\mathcal{O}(KR \ln(KR))$.

\subsubsection{PrivateDFL vs. Black-Box Noise Accumulation}
PrivateDFL maintains logarithmic noise scaling with respect to the total number of iterations:
\begin{equation}
\sigma^2_{\text{PrivateDFL}} = \mathcal{O}\big(\ln(KR)\big),
\end{equation}
whereas the black-box composition suffers from super-linear noise accumulation:
\begin{equation}
\sigma^2_{\text{Black-Box}} = \mathcal{O}\big(KR \ln(KR)\big).
\end{equation}
Thus, cumulative noise tracking is essential for preserving accuracy in decentralized differentially private learning.

\begin{algorithm}
\caption{PrivateDFL: Private Decentralized Federated Learning}
\label{alg:privatedfl}
\begin{algorithmic}[1]
\setlength{\itemsep}{0.1em}
\setlength{\parskip}{0.1em}
\State \textbf{Input} $\{\Vec{F}_{k}\}_{k=1}^{K}$ \Comment{Training samples from $K$ clients}
\State \textbf{Output} $\{\tilde{\Vec{C}}_{s,K}^{\,R}\}_{s=1}^{S}$ \Comment{Final secure HD model after $R$ rounds (state at client $K$)}
\State \textbf{Encoding} $\Vec{H}=(\cos(\Vec{F}\!\cdot\!\Vec{B}_d))_{d=1}^{D}$, $\{\Vec{B}_d\}_{d=1}^{D}\!\sim\!\mathcal{N}(0,1)$
\State \textbf{ClassHV} $\Vec{C}_s=\sum \Vec{H}^{s},\; s=1{:}S$
\State \textbf{Retrain} $\Vec{C}_{s'}\!\leftarrow\!\Vec{C}_{s'}-\Vec{H}^{s}$, $\Vec{C}_{s}\!\leftarrow\!\Vec{C}_{s}+\Vec{H}^{s}$
\For{$r=1{:}R$}
  \For{$k=1{:}K$}
    \If{$r=1 \land k=1$}
      \State $\vec{\Gamma}_{1}^{1} \sim \mathcal{N}\!\left(0, \tfrac{2D}{\epsilon^{2}} \ln\!\left[\tfrac{1.25N}{\delta_{0}}\right]\right)$
      \State $\{\tilde{\Vec{C}}_{s,1}^{\,1}\}_{s=1}^{S} \leftarrow \text{ClassHV}(\text{Encoding}(\Vec{F}_{1})) + \vec{\Gamma}_{1}^{1}$
    \ElsIf{$r=1 \land k\neq 1$}
      \State $\vec{\Gamma}_{k}^{r} \sim \mathcal{N}\!\left(0, \tfrac{2D}{\epsilon^{2}} \ln\!\left[\tfrac{K(r-1)+k}{K(r-1)+k-1}\right]\right)$
      \State $\begin{aligned}[t]
       \{\tilde{\Vec{C}}_{s,k}^{\,1}\}_{s=1}^{S} \leftarrow{}& \{\tilde{\Vec{C}}_{s,k-1}^{\,1}\}_{s=1}^{S} + \text{ClassHV}(\text{Encoding}(\Vec{F}_{k})) + \vec{\Gamma}_{k}^{1}
      \end{aligned}$
    \Else
      \State $\vec{\Gamma}_{k}^{r} \sim \mathcal{N}\!\left(0, \tfrac{2D}{\epsilon^{2}} \ln\!\left[\tfrac{K(r-1)+k}{K(r-1)+k-1}\right]\right)$
      \State $\begin{aligned}[t]
        \{\tilde{\Vec{C}}_{s,k}^{\,r}\}_{s=1}^{S} \leftarrow{}&
          \text{Retrain}\!\big(\{\tilde{\Vec{C}}_{s,k-1}^{\,r}\}_{s=1}^{S}, \text{Encoding}(\Vec{F}_{k})\big) + \vec{\Gamma}_{k}^{r}
      \end{aligned}$
    \EndIf
  \EndFor
  \If{$r<R$}
    \State $\{\tilde{\Vec{C}}_{s,1}^{\,r+1}\}_{s=1}^{S} \leftarrow \{\tilde{\Vec{C}}_{s,K}^{\,r}\}_{s=1}^{S}$
  \EndIf
\EndFor
\State \Return $\{\tilde{\Vec{C}}_{s,K}^{\,R}\}_{s=1}^{S}$
\end{algorithmic}
\end{algorithm}

\section{Experimental Design}
\label{sec:Experimental Design}

We evaluate PrivateDFL on three benchmark datasets spanning distinct sensing modalities: MNIST (image)~\cite{deng2012mnist}, ISOLET (speech)~\cite{fanty1990spoken}, and UCI-HAR (wearable sensors)~\cite{asuncion2007uci}. This selection enables testing across heterogeneous input domains. Both IID and non-IID settings are considered. In the IID setting, all training samples are pooled and evenly partitioned across the $K$ clients without regard to label, so each client observes approximately the same class distribution. In the non-IID setting, each client is assigned exactly two classes and receives only samples from those two classes, so the number of clients is determined by the number of classes in each dataset: five clients for MNIST, thirteen for ISOLET, and three for UCI-HAR. For example, the MNIST clients hold $\{0,1\}$, $\{2,3\}$, and so on. This partition reflects realistic deployments where each client observes only a narrow slice of the label space.

The evaluation proceeds in three steps. First, we conduct a sensitivity analysis to study how the privacy budget $\epsilon$, privacy-loss parameter $\delta_0$, hypervector dimensionality $D$, the number of clients $K$, and the number of training samples per client affect accuracy under both IID and non-IID partitions. Because the number of clients in the non-IID setting is fixed by the class count, the client-count sweep is conducted under the IID partition, where we vary $K$ while keeping the total dataset fixed to isolate the effect of data fragmentation.

\begin{figure*}[!t]
\centering
\subfloat[]{\includegraphics[width=0.33\textwidth]{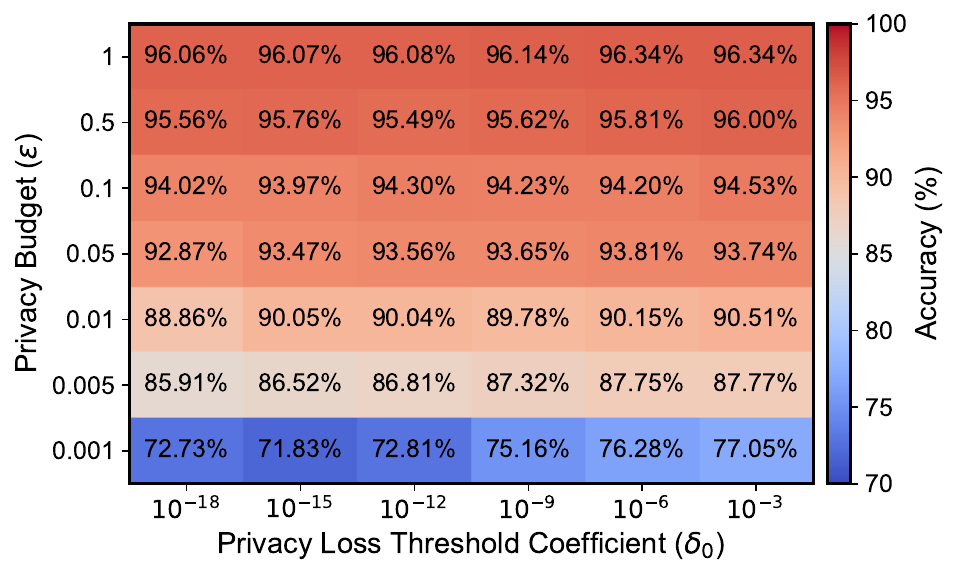}\label{fig:eps_delta_iid_mnist}} \hfill
\subfloat[]{\includegraphics[width=0.33\textwidth]{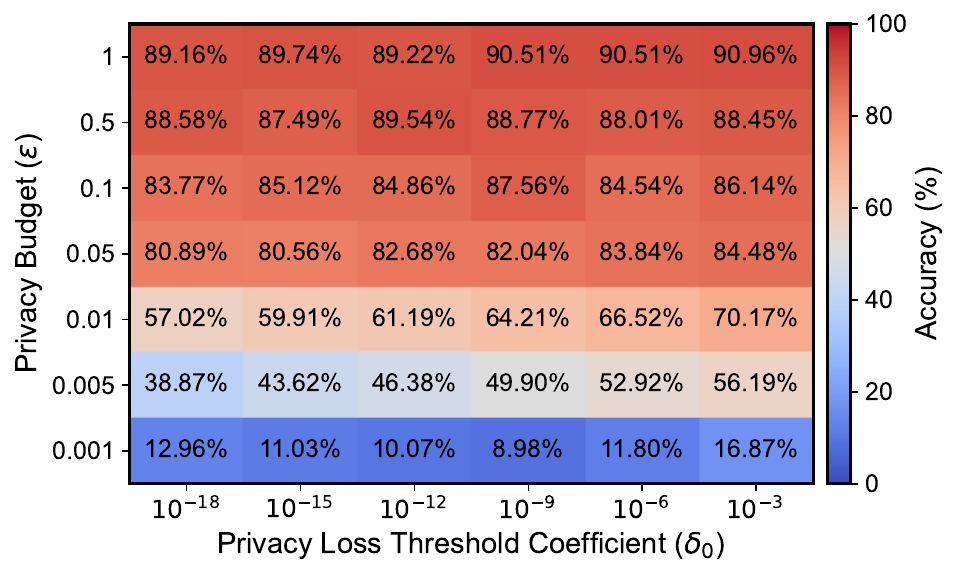}\label{fig:eps_delta_iid_isolet}} \hfill
\subfloat[]{\includegraphics[width=0.33\textwidth]{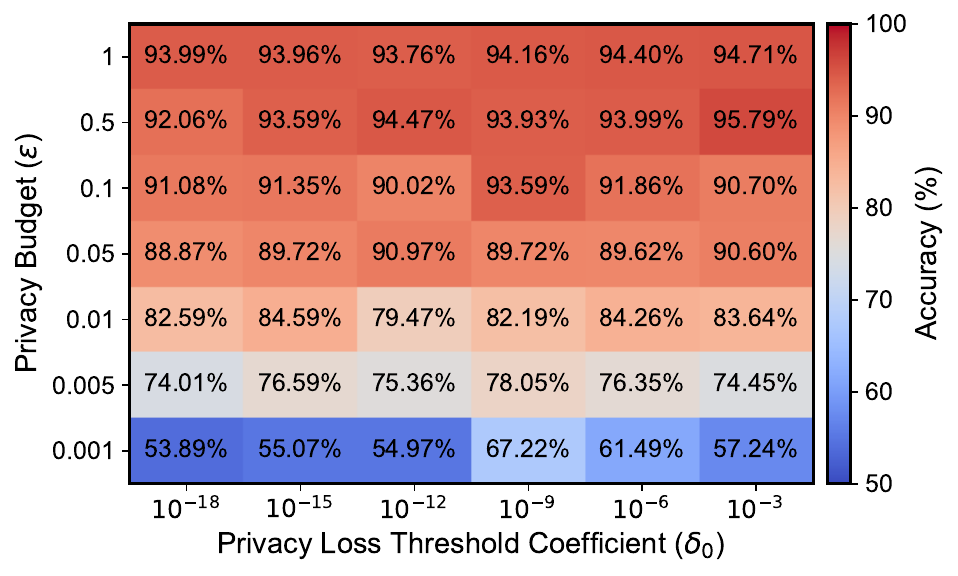}\label{fig:eps_delta_iid_ucihar}}\\
\subfloat[]{\includegraphics[width=0.33\textwidth]{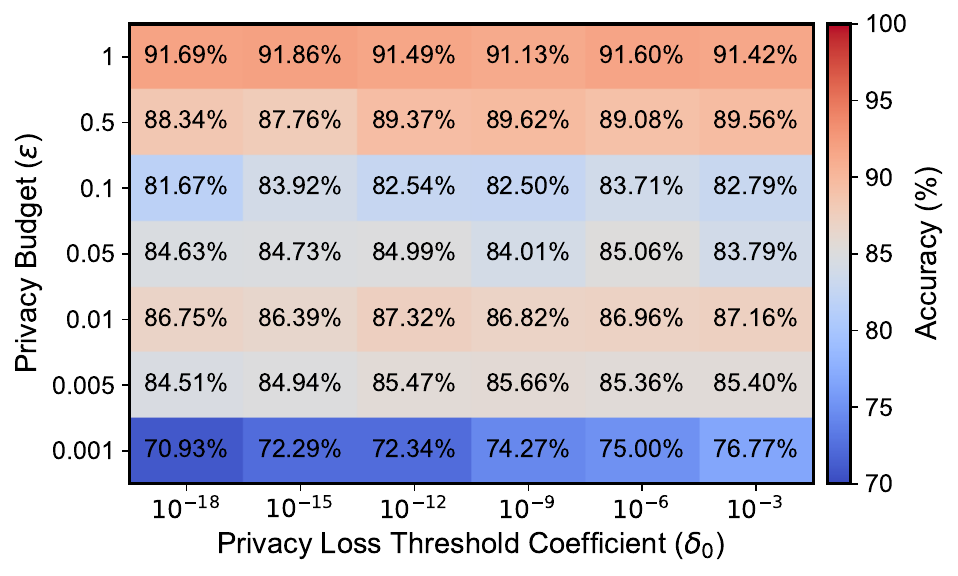}\label{fig:eps_delta_non_mnist}} \hfill
\subfloat[]{\includegraphics[width=0.33\textwidth]{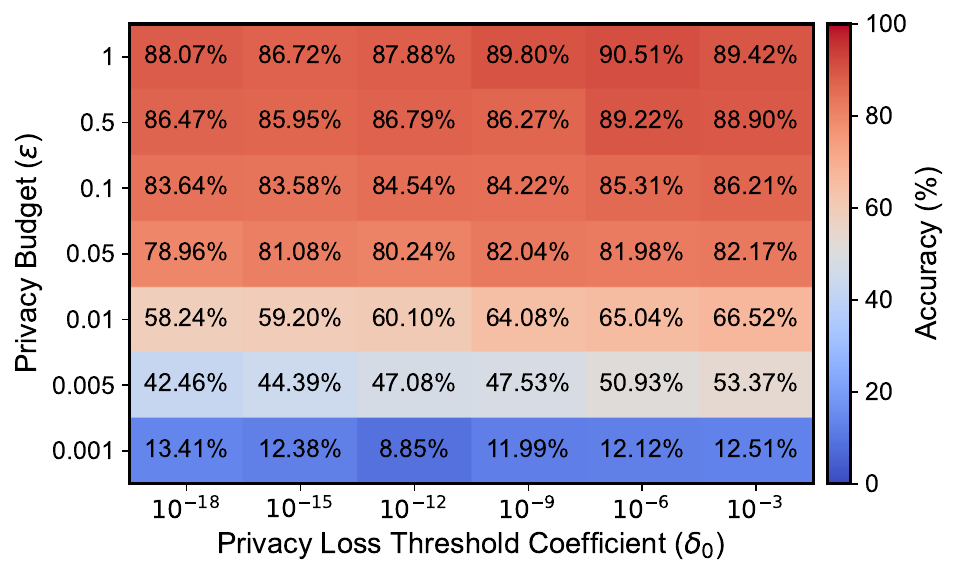}\label{fig:eps_delta_non_isolet}} \hfill
\subfloat[]{\includegraphics[width=0.33\textwidth]{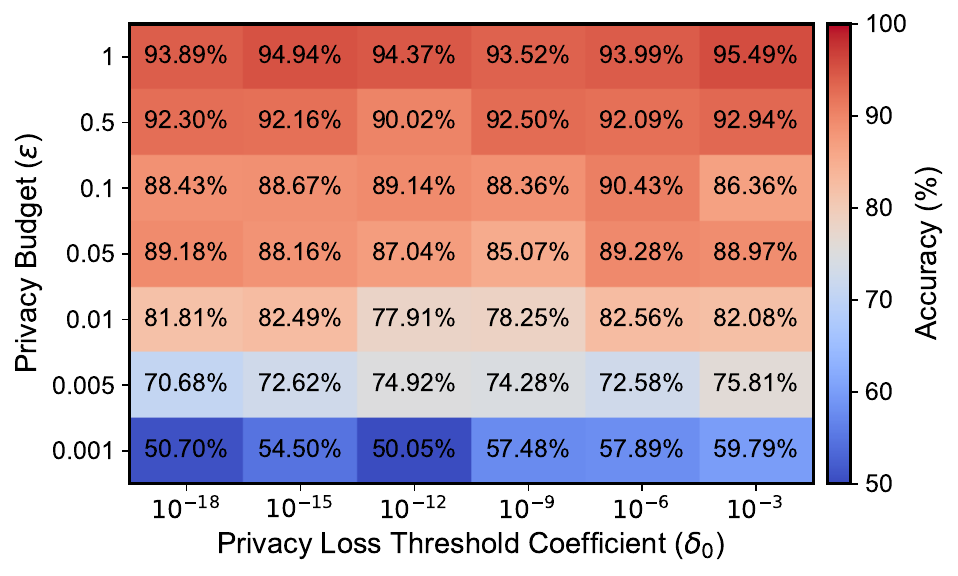}\label{fig:eps_delta_non_ucihar}}
\caption{Accuracy of PrivateDFL under varying privacy parameters, including the privacy budget ($\varepsilon$) and privacy loss coefficient ($\delta_{0}$), for both IID and non-IID data distributions. Panels (a)--(c) report results on MNIST, ISOLET, and UCI-HAR under IID partitioning, while panels (d)--(f) show the corresponding outcomes under non-IID splits.}
\label{fig:eps_delta_sensitivity}
\end{figure*}

\begin{figure}[!t]
\centering
\includegraphics[width=0.4\textwidth]{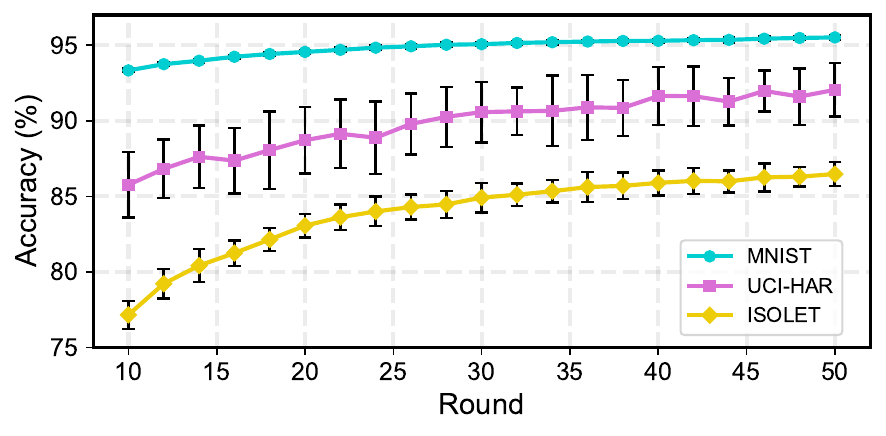}
\caption{Accuracy of PrivateDFL across communication rounds under IID partitioning. For each dataset (MNIST, ISOLET, UCI-HAR), the curve reports the mean accuracy over client counts \(K \in \{10,\ldots,1000\}\), and the shaded region denotes the corresponding \(\pm 1\) standard deviation.}
\label{fig:IID_nclients_acc}
\end{figure}

Second, we benchmark PrivateDFL against state-of-the-art Transformer and deep learning models trained with DP in centralized settings. These baselines access the full dataset directly and therefore represent an upper bound on accuracy and latency without federated constraints, so surpassing them implies an advantage over their decentralized counterparts as well. We implement DP-SGD and RDP using Opacus~\cite{yousefpour2021opacus}, tuning each baseline's DP hyperparameters, including the clipping norm, learning rate, and noise multiplier, for the target budget. MNIST baselines include Vision Transformer, ResNet50, GoogLeNet, and AlexNet; ISOLET baselines include Transformer, Convolutional Neural Network (CNN), Recurrent Neural Network (RNN), and Temporal Convolutional Network (TCN); and UCI-HAR baselines include TabTransformer, TabNet, Deep Neural Network (DNN), and Deep and Cross Network (DCN).

All benchmark comparisons use a fixed privacy budget of $\epsilon=0.4$ and privacy-loss coefficient $\delta_0=0.001$, while the sensitivity analysis additionally sweeps a wide range of both parameters. The benchmark point is a deliberately strict operating regime: a small $\epsilon$ injects large noise and stresses every method at a high privacy level, which is precisely the regime in which the privacy--utility tradeoff matters most. Under such tight budgets, gradient-based deep models are disproportionately affected, particularly on sequential inputs, whereas PrivateDFL's noise accounting injects only the minimal incremental noise required to meet the same budget and thereby preserves accuracy. We report all methods in terms of accuracy, training time, inference latency, and energy consumption.

All experiments, for both PrivateDFL and the baselines, are run on a single NVIDIA GeForce RTX 3060 with identical hardware and software. Energy consumption is measured with the Zeus monitor, and training and inference times use CUDA-synchronized wall-clock timing, applied identically across all methods.

\begin{figure*}[!t]
\centering
\subfloat[]{\includegraphics[width=0.33\textwidth]{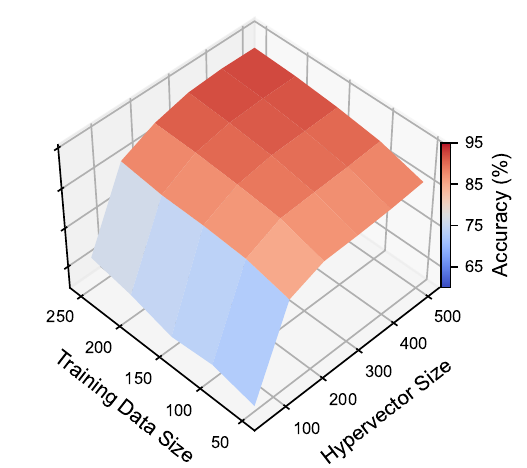}\label{fig:IID_Nsamples_D_acc_mnist}} \hfill
\subfloat[]{\includegraphics[width=0.33\textwidth]{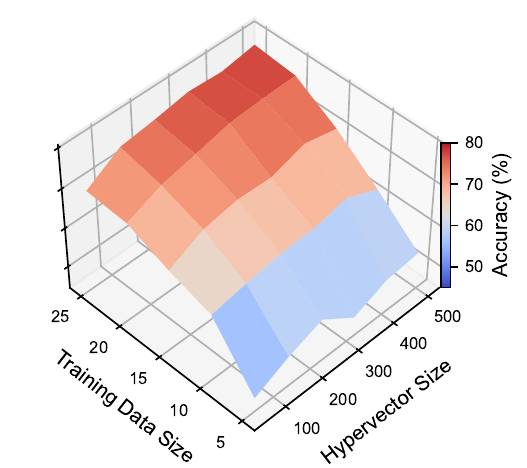}\label{fig:IID_Nsamples_D_acc_isolet}} \hfill
\subfloat[]{\includegraphics[width=0.33\textwidth]{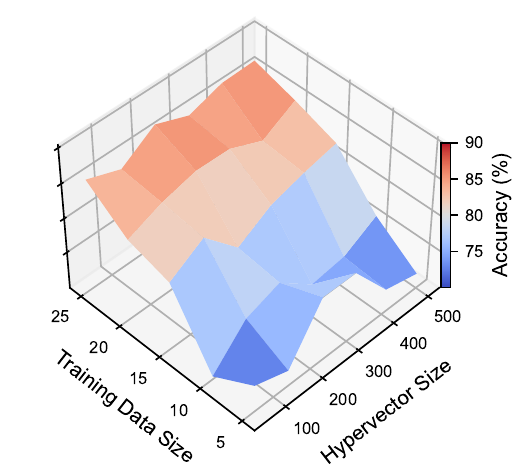}\label{fig:IID_Nsamples_D_acc_ucihar}}
\caption{Effect of hypervector dimensionality and the number of training samples per client on model accuracy under IID partitioning. Panels (a)--(c) report results for MNIST, ISOLET, and UCI-HAR, respectively.}
\label{fig:Nsamples_D_sensitivity}
\end{figure*}

\begin{figure}[!t]
\centering
\includegraphics[width=0.4\textwidth]{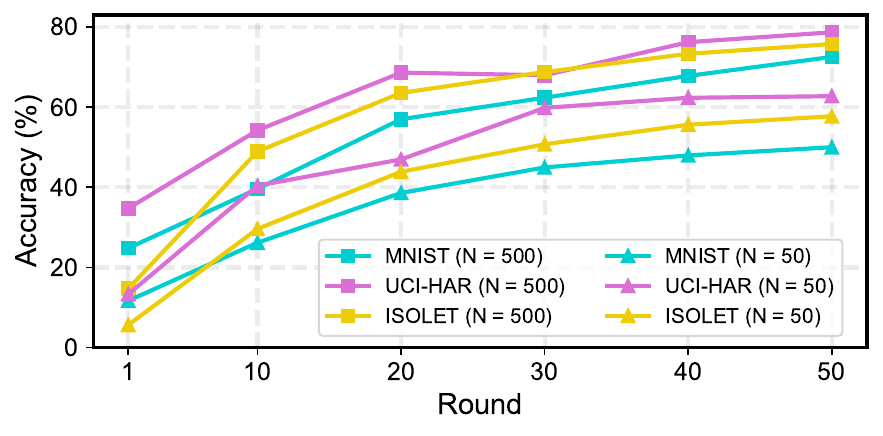}
\caption{Accuracy of PrivateDFL across communication rounds under non-IID partitioning, shown for varying numbers of training samples per client.}
\label{fig:nonIID_trainsamples_acc}
\end{figure}

\begin{figure}[!t]
\centering
\includegraphics[width=0.4\textwidth]{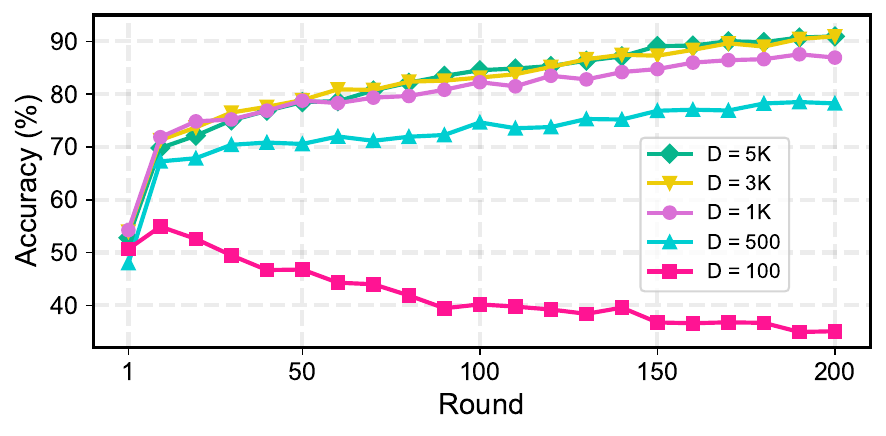}
\caption{PrivateDFL accuracy across communication rounds for varying hypervector dimensionalities ($D$) under non-IID MNIST partitioning.}
\label{fig:mnist_non_iid_D_acc}
\end{figure}

\section{Experimental Results}
\label{sec:Experimental Results}

\subsection{Sensitivity Analysis}
\label{sec:Sensitivity Analysis}

In this subsection, we examine how key hyperparameters influence the performance of PrivateDFL. The parameters of interest include the privacy budget $\epsilon$, the privacy loss coefficient $\delta_{0}$, the number of clients $K$, the hypervector dimensionality $D$, and the number of training samples per client. We begin by analyzing the effect of the privacy parameters under IID conditions.

Figure~\ref{fig:eps_delta_iid_mnist} reports the MNIST accuracy under the IID setting with $100$ clients. With a large privacy budget ($\epsilon=1$, $\delta_{0}=10^{-3}$), the injected noise is minimal and PrivateDFL attains $96.34\%$. Fixing $\delta_{0}=10^{-9}$ and decreasing $\epsilon$ increases the noise and monotonically reduces accuracy, from $96.14\%$ at $\epsilon=1$ down to $75.16\%$ at $\epsilon=0.001$. In contrast, varying $\delta_{0}$ has a marginal effect: at $\epsilon=0.1$, tightening $\delta_{0}$ from $10^{-3}$ to $10^{-18}$ changes accuracy by only about $0.56$ percentage points, confirming that $\epsilon$ dominates the privacy--utility tradeoff. ISOLET and UCI-HAR follow the same pattern under IID partitions, as shown in Figures~\ref{fig:eps_delta_iid_isolet} and~\ref{fig:eps_delta_iid_ucihar}.

Figures~\ref{fig:eps_delta_non_mnist}, \ref{fig:eps_delta_non_isolet}, and \ref{fig:eps_delta_non_ucihar} extend this analysis to the non-IID setting, where each client holds only two classes and the number of clients is fixed by the class count (five for MNIST, thirteen for ISOLET, and three for UCI-HAR). The overall trends remain consistent, though the effect is more pronounced for MNIST due to greater heterogeneity: at $\epsilon=0.5$ and $\delta_{0}=10^{-6}$, MNIST accuracy declines from $95.81\%$ under IID to $89.08\%$ under non-IID, while ISOLET and UCI-HAR change by less than two percentage points.

Figure~\ref{fig:IID_nclients_acc} examines the effect of client count under IID partitioning across communication rounds. Because the non-IID partition fixes the client count by the number of classes, this study is conducted under the IID setting, varying $K$ from $10$ to $1000$ while keeping the total training set fixed so that changes arise from partitioning rather than data volume. The variability across client counts remains small for MNIST (below $1.5\%$), moderate for ISOLET (up to $5.31\%$), and larger for UCI-HAR (up to $7.37\%$), and stays nearly constant across rounds, indicating that the primary source of uncertainty is the intrinsic task difficulty rather than the progression of rounds.

Figure~\ref{fig:Nsamples_D_sensitivity} illustrates the joint effect of hypervector dimensionality and the number of training samples per client under IID partitioning, for MNIST, ISOLET, and UCI-HAR in Figures~\ref{fig:IID_Nsamples_D_acc_mnist}, \ref{fig:IID_Nsamples_D_acc_isolet}, and \ref{fig:IID_Nsamples_D_acc_ucihar}. For MNIST, accuracy is driven primarily by dimensionality: with 200 samples per client it rises from $65.35\%$ at $D=50$ to $91.86\%$ at $D=500$, whereas increasing the sample count at fixed $D=300$ yields a smaller gain. ISOLET and UCI-HAR show the opposite trend, where the number of training samples matters more than dimensionality; for ISOLET at $D=500$, accuracy improves from $53.56\%$ to $77.99\%$ as samples grow from 5 to 25, while raising $D$ at a fixed sample count yields only modest gains. These results indicate that dimensionality dominates for image data, whereas sample count dominates for speech and sensor data.

\begin{figure*}[!t]
\centering
\subfloat[]{\includegraphics[width=0.33\textwidth]{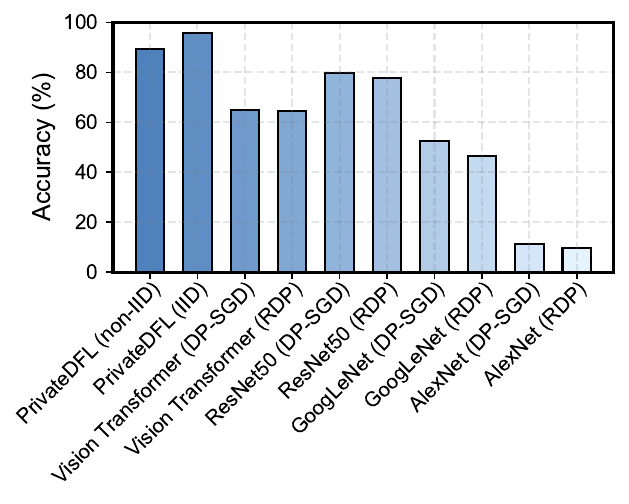}\label{fig:mnist_bench_acc}} \hfill
\subfloat[]{\includegraphics[width=0.33\textwidth]{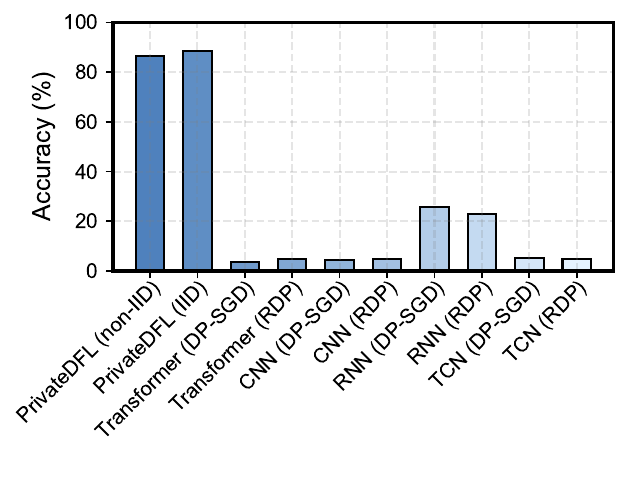}\label{fig:isolet_bench_acc}} \hfill
\subfloat[]{\includegraphics[width=0.33\textwidth]{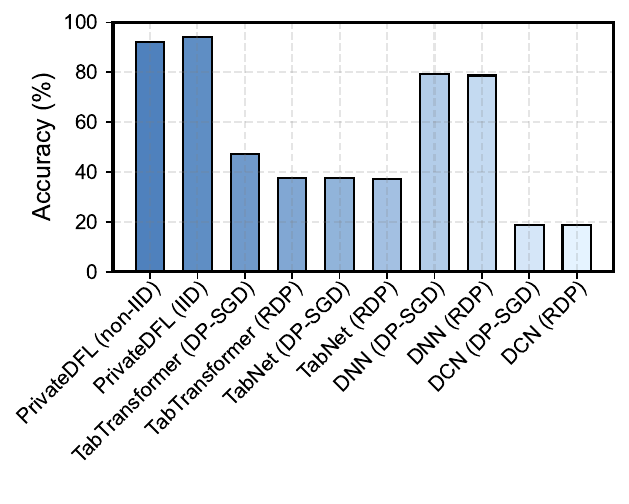}\label{fig:ucihar_bench_acc}}
\caption{Accuracy comparison between PrivateDFL and differentially private baseline models across (a) MNIST, (b) ISOLET, and (c) UCI-HAR datasets.}
\label{fig:bench_accuracy}
\end{figure*}

Extending to the non-IID setting, Figure~\ref{fig:nonIID_trainsamples_acc} reports accuracy across rounds while varying the number of samples per client. Additional training samples consistently improve performance at all rounds; for MNIST, the round-50 accuracy rises from $50.00\%$ with 50 samples per client to $72.53\%$ with more data, and similar gains appear for ISOLET and UCI-HAR, whose round-50 accuracy increases from $57.72\%$ and $62.74\%$ to $75.75\%$ and $78.69\%$ as samples grow to 500 per client.

Finally, Figure~\ref{fig:mnist_non_iid_D_acc} evaluates dimensionality under non-IID MNIST partitions. When $D$ is small ($D=100$), the model lacks capacity and the DP noise dominates, so accuracy declines over rounds to $35.10\%$ by round 200. Increasing $D$ reverses this: $D=500$ reaches $78.25\%$ at round 200, and $D=1000$ and $D=3000$ reach $86.93\%$ and $90.95\%$. Beyond this range the benefit saturates, as larger $D$ also amplifies the effective DP noise; $D=5000$ reaches $90.94\%$, nearly identical to $D=3000$.

\subsection{Benchmark}
\label{sec:Benchmark}

Figure~\ref{fig:bench_accuracy} compares PrivateDFL with differentially private baselines across the three datasets under both partitions. All baselines are trained centrally with tuned DP hyperparameters and therefore represent an upper bound relative to their decentralized counterparts. Across all datasets and distributions, PrivateDFL consistently achieves the highest accuracy while satisfying the same privacy budgets.

For MNIST (Figure~\ref{fig:mnist_bench_acc}), PrivateDFL attains $95.74\%$ under IID and $89.38\%$ under non-IID. The DP-trained deep baselines fall well below this, ranging from $9.51\%$ (AlexNet) to $79.59\%$ (ResNet50). For ISOLET (Figure~\ref{fig:isolet_bench_acc}), PrivateDFL reaches $88.45\%$ (IID) and $86.66\%$ (non-IID), while Transformer and CNN baselines remain between $3.78\%$ and $5.26\%$, near the chance level for this $26$-class task, and the strongest DP baseline (RNN) reaches only $25.59\%$. For UCI-HAR (Figure~\ref{fig:ucihar_bench_acc}), PrivateDFL achieves $94.30\%$ (IID) and $92.33\%$ (non-IID), against a best tabular baseline (DNN) of $79.47\%$. Under this strict privacy budget, the large noise required by DP-SGD and RDP overwhelms the gradient-based models, whereas PrivateDFL injects only the minimal incremental noise needed to meet the same budget and thereby preserves accuracy across image, speech, and sensor modalities.

\begin{figure*}[!t]
\centering
\subfloat[]{\includegraphics[width=0.33\textwidth]{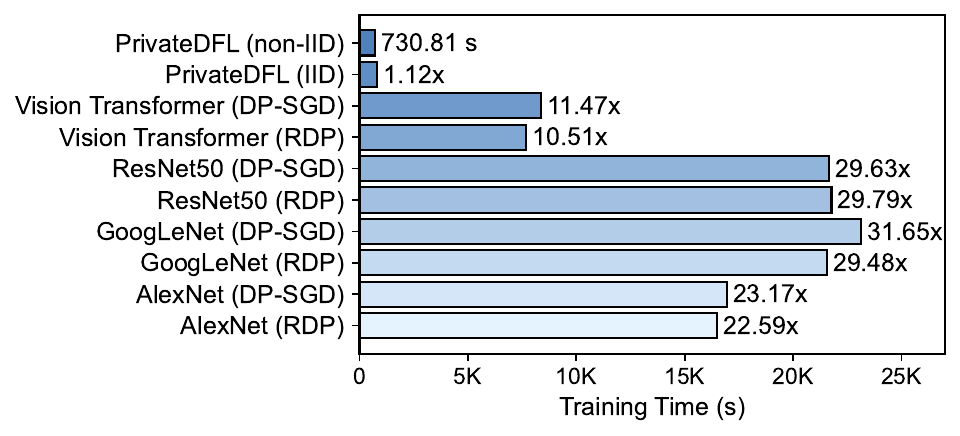}\label{fig:mnist_bench_traintime}} \hfill
\subfloat[]{\includegraphics[width=0.33\textwidth]{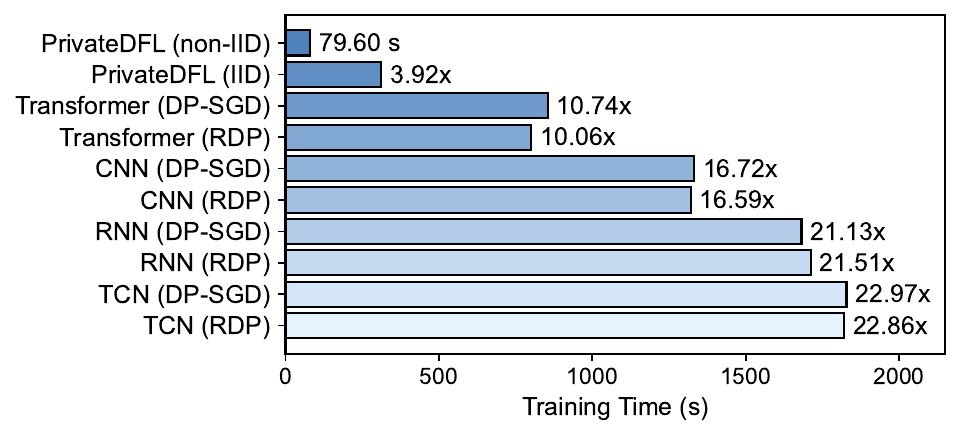}\label{fig:isolet_bench_traintime}} \hfill
\subfloat[]{\includegraphics[width=0.33\textwidth]{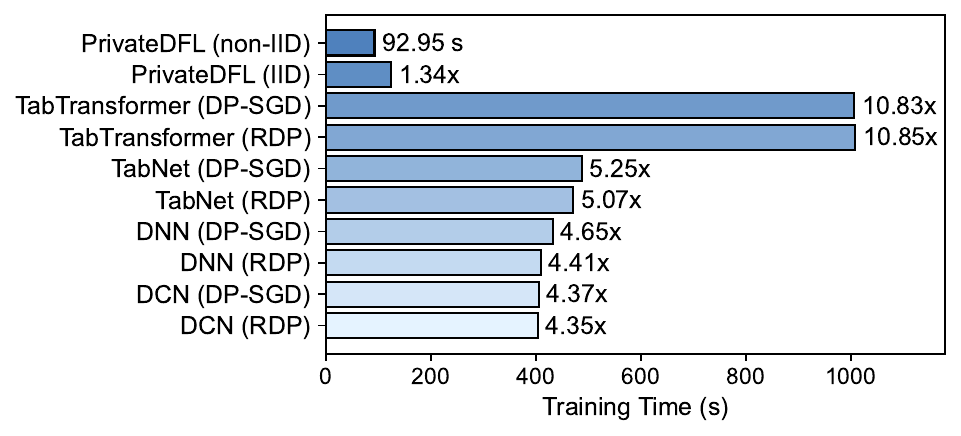}\label{fig:ucihar_bench_traintime}}\\
\subfloat[]{\includegraphics[width=0.33\textwidth]{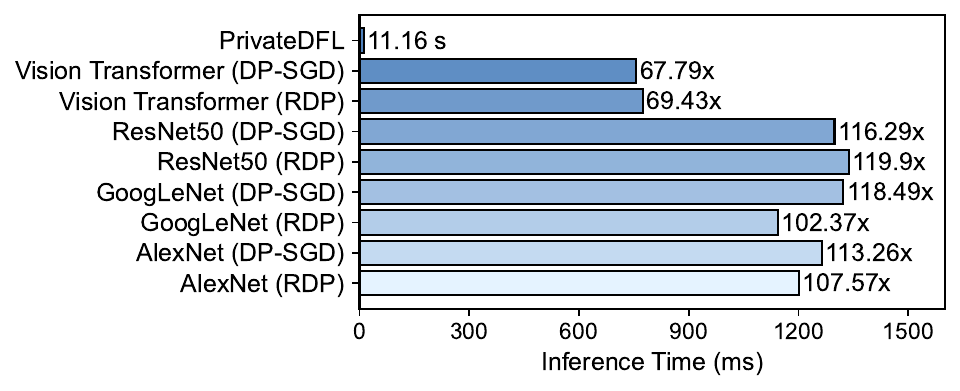}\label{fig:mnist_bench_infertime}} \hfill
\subfloat[]{\includegraphics[width=0.33\textwidth]{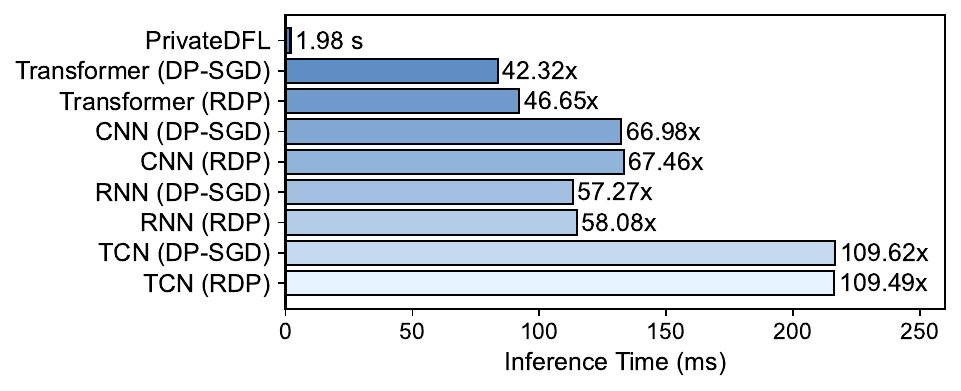}\label{fig:isolet_bench_infertime}} \hfill
\subfloat[]{\includegraphics[width=0.33\textwidth]{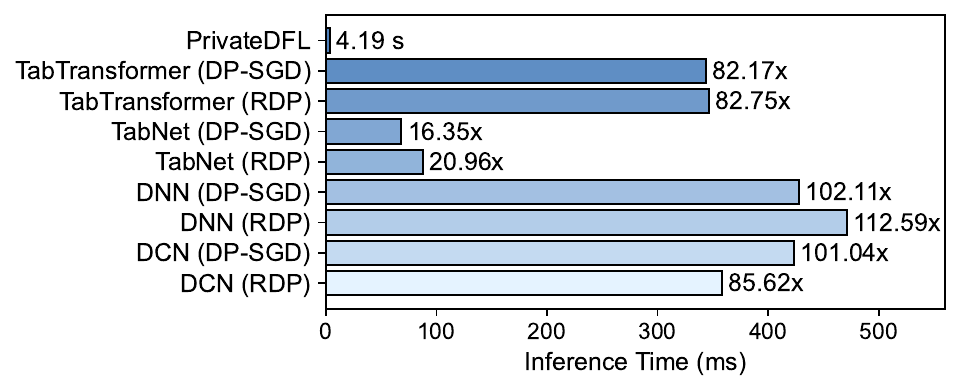}\label{fig:ucihar_bench_infertime}}
\caption{Comparison of runtime performance for PrivateDFL and DP baseline models. The top row shows training time and the bottom row shows inference latency for (a,d) MNIST, (b,e) ISOLET, and (c,f) UCI-HAR.}
\label{fig:bench_time}
\end{figure*}

Figure~\ref{fig:bench_time} compares training time and inference latency. PrivateDFL completes MNIST training in $820.97$ seconds (IID) and $730.81$ seconds (non-IID), whereas the deep baselines require between roughly $7{,}600$ and $23{,}000$ seconds. The gap is even larger at inference: PrivateDFL classifies in $11.16$ milliseconds on MNIST, $1.98$ on ISOLET, and $4.19$ on UCI-HAR, while the deep and sequential baselines range from tens to over a thousand milliseconds. These results show that PrivateDFL attains higher accuracy at a small fraction of the computational cost.

\begin{figure*}[!t]
\centering
\subfloat[]{\includegraphics[width=0.33\textwidth]{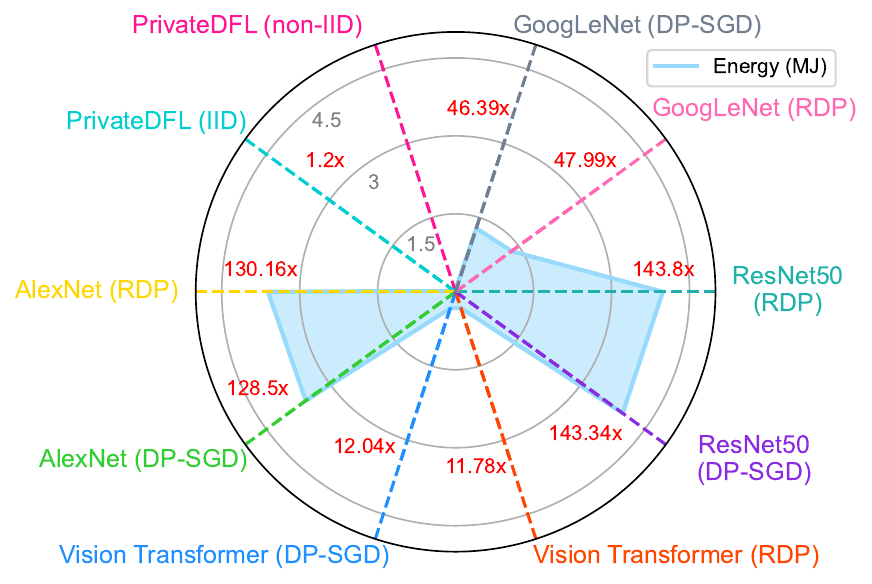}\label{fig:mnist_bench_energy}} \hfill
\subfloat[]{\includegraphics[width=0.33\textwidth]{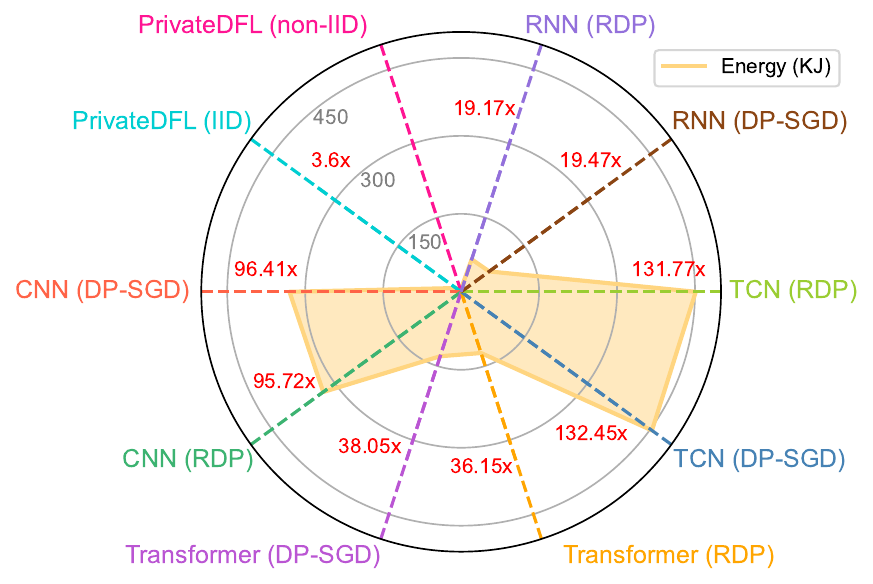}\label{fig:isolet_bench_energy}} \hfill
\subfloat[]{\includegraphics[width=0.33\textwidth]{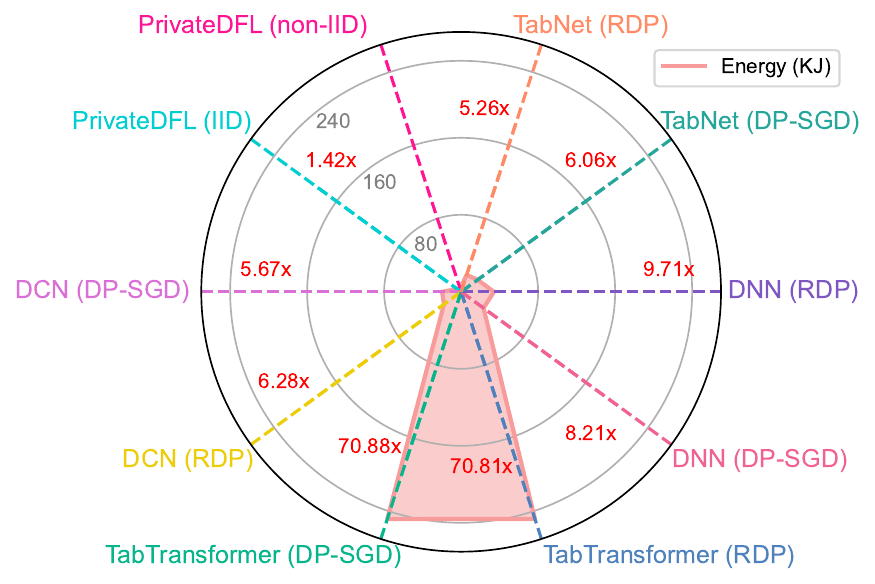}\label{fig:ucihar_bench_energy}}
\caption{Energy consumption comparison between PrivateDFL and differentially private baseline models across the three benchmark datasets. Subfigures (a)--(c) correspond to MNIST, ISOLET, and UCI-HAR, respectively.}
\label{fig:bench_energy}
\end{figure*}

Figure~\ref{fig:bench_energy} reports energy consumption, which follows the accuracy and timing trends. On MNIST, PrivateDFL consumes only $0.03$ MJ, at least $11\times$ less than the most efficient deep baseline (Vision Transformer) and up to roughly $130\times$ less than ResNet50. Similar margins hold on ISOLET ($12.32$ kJ, up to $130\times$ less than TCN) and UCI-HAR ($4.96$ kJ, the lowest among all evaluated models). Overall, PrivateDFL consistently achieves high utility while substantially lowering energy consumption across all three modalities, underscoring its suitability for resource-constrained deployment.

\section{Discussion and Limitations}
\label{sec:Discussion and Limitations}

The results across image, speech, and sensor modalities show that the advantage of PrivateDFL arises from its noise accounting rather than from the choice of classifier. Under the strict privacy budget used for benchmarking, gradient-based deep models must inject large perturbations at every step, which drives several of them close to chance. PrivateDFL instead tracks the cumulative perturbation already present in the shared model and injects only the minimal incremental noise required to meet the same budget, so accuracy is preserved. The effect of this accounting can be isolated by comparing the cumulative noise with and without tracking. As derived in Section~\ref{sec:RM Cumulative Noise}, the tracked variance grows only logarithmically in the number of clients and rounds, whereas the untracked variant grows super-linearly, scaling as their product times its logarithm. The utility consequence of this difference is visible in Figure~\ref{fig:eps_delta_sensitivity}, where accuracy falls monotonically as the injected noise increases. Combining the two observations, the factorial growth of the untracked scheme drives it into the low-accuracy regime, while the logarithmic growth of PrivateDFL keeps it in the high-accuracy regime. The accountant, rather than the hyperdimensional backbone, is therefore the source of the privacy--utility advantage under strict privacy.

The characteristics of PrivateDFL directly support applications in domains where data cannot be centralized because of confidentiality, regulation, or ownership. Many real-world deployments involve sensitive information, including patient health records, financial transactions, and government-managed personal identifiers. In such environments, strong privacy guarantees and transparent accounting are required for any collaborative learning system. PrivateDFL guarantees that each participant's contribution remains indistinguishable while tracking cumulative noise explicitly across all rounds, which avoids the severe utility loss that affects deep models trained with DP. Industrial environments, such as aerospace, automotive, and energy systems, also involve proprietary operational data. These settings require secure collaborative learning for tasks such as real-time defect detection, anomaly monitoring, and predictive maintenance. PrivateDFL is well-suited for these scenarios, since it maintains high accuracy under strict privacy constraints while operating efficiently on distributed devices.

Several limitations remain. First, the ring topology updates clients sequentially rather than in parallel, so training time grows with the number of clients, and at very large client counts this serialization becomes a computational bottleneck. This is the cost of the decentralized, server-free design that yields the accuracy of PrivateDFL under strict privacy. In centralized DP-FL, parallel aggregation keeps training time low but accuracy under strict budgets is often limited; for instance, our prior centralized framework~\cite{piran2025privacy} reaches under $71\%$ on a three-class task even at a weak budget, whereas PrivateDFL exceeds $90\%$ under much stronger privacy. A promising direction is a hybrid design that recovers parallelism for large client counts while preserving the accuracy benefits of incremental noise accounting. Second, the current framework assumes synchronous ring-style communication and reliable message passing, and extending the accounting to asynchronous schedules, dynamic topologies, and intermittent participation remains open. Third, PrivateDFL uses a uniform privacy target for all clients, and adaptive budgets that reflect local data sensitivity while preserving global guarantees are left to future work. Finally, the present threat model considers an honest-but-curious adversary, so robustness to active adversarial behavior, including data poisoning, backdoor insertion, collusion, and free-riding, requires deeper study.

\section{Conclusion and Future Work}
\label{sec:Conclusions and Future Work}

This work introduced PrivateDFL, a privacy-preserving decentralized learning framework that integrates HD with differentially private noise scheduling and transparent accounting. The framework provides three core advantages. First, HD classification offers an interpretable structure based on class prototypes and explicit update rules. Second, privacy is formally guaranteed through calibrated Gaussian noise whose magnitude is analytically derived for every client and communication round, so that every model exchanged between clients satisfies the target budget. Third, the lightweight HD operations make PrivateDFL practical on resource-constrained devices due to low computational cost, low inference latency, and minimal communication overhead.

Although baseline models are trained centrally and therefore operate under more favorable conditions, PrivateDFL consistently matches or exceeds their performance under equal privacy budgets. Across image, speech, and sensor modalities, the framework achieves $95.74\%$ on MNIST, $88.45\%$ on ISOLET, and $94.30\%$ on UCI-HAR under IID settings, with similar patterns under non-IID. These improvements coincide with much lower training time, inference latency, and energy usage, which makes PrivateDFL an effective and efficient option for decentralized learning.

Future work will extend the noise accounting to asynchronous schedules, dynamic topologies, and intermittent participation, and will investigate hybrid decentralized-centralized designs that recover parallelism at large client counts. Further directions include adaptive per-client privacy budgets that reflect local data sensitivity while preserving global guarantees, and strengthening the protocol against active adversaries through secure aggregation, lightweight homomorphic encryption, or verifiable update mechanisms. In summary, PrivateDFL demonstrates that interpretable hyperdimensional representations, decentralized learning, and explainable DP can be combined to deliver high accuracy, strong privacy, low latency, and low energy consumption, supporting its use as a practical and scalable solution for trustworthy privacy-preserving learning in distributed and sensitive environments.

\section*{Acknowledgment}
This work was supported by the National Science Foundation, United States, under Grant 2434519. The code used in this study is openly available at \href{https://github.com/FardinJalilPiran/PrivateDFL}{https://github.com/FardinJalilPiran/PrivateDFL}.

\bibliographystyle{unsrt}
\bibliography{references}

\end{document}